# Photon echoes for a system of large negative spin and few photons

Michael T. Tavis

## Abstract

**Persistent photon echoes are seen for the case of a large number of two-level molecules (TLMs) prepared initially in the all-down state (large negative spin) interacting with a photon distribution with a small mean photon number in a lossless cavity. This case is interesting since 1) it has not been significantly addressed in the past; 2) the characteristic times associated with revival are not what is seen for the more frequently addressed problem of a few TLMs interacting with a photon distribution with a large mean photon number; 3) long after the echoes die out, they re-emerge completely at a much later time; and 4) the existence of echoes is not predicated on the initial photon distribution being the coherent state or Glauber distribution. Entropy, entanglement, and the Q function are considered. It is found that disentanglement only occurs at the revival times as evidenced by the entropy going to 0 and the Q function returning to the value seen at time=0. Comparison to the more normally seen results for large mean photon number and small numbers of TLMs are discussed. Finally this paper acts as a general reference for future refereed publications.**

## 1 Introduction

The purpose of this paper is to describe the behavior of photon echoes which occur for a system of many two-level molecules (TLMs) all prepared in the ground state (large negative spin) interacting with a photon field with small mean photon number (SMPN) in a closed cavity. This particular example in a crowded active field of research has been overlooked. The interaction of TLMs interacting with a field in a closed cavity is well known and has been covered extensively, as will be seen in the next section. An examination of the time behavior of such a system is undertaken in this paper from the perspective of the field rather than the TLMs as is done for most articles in this area of research. The characteristic times associated with the echo behavior does not have the same behavior seen for small numbers of TLMs interacting with strong fields. To emphasize this observation:1) The photon echo revival time found in this paper is given by $\tau_R = 4\pi\sqrt{N + \Delta}/\gamma$, where N is the number of TLMs and $\Delta$ is the measure of detuning between photon energy and the energy separation for the TLM. $\gamma$ is the coupling constant between TLMs and photons. This is in contrast to $2\pi\sqrt{\bar{n} + 1 + \Delta}/\gamma$, where $\bar{n}$ is the mean photon number which was found for few TLMs and a large mean photon number (LMPN); 2) a complete set of revival echoes is seen for a time proportional to the number of TLMs times $\tau_R$ and again for later times after the initial echo set has died away by $\sqrt{N}\ \tau_R$. This is not a behavior seen for small numbers of TLMs and LMPN; and 3) echo behavior is independent of the initial photon distribution, an example being the thermal photon distribution, which is contrary to the behavior seen for the all spin-up case.

A somewhat limited history of research leading to this work is found in the next section (Background). A complete accounting of all important research would be impossible within the limits of this paper. In the third section the characteristic echo behavior (A discussion of characteristic time) for the two regimes is considered. The first subsection, (Characteristic times for few TLMs and LMPN), is somewhat of a review for the well-known case of small numbers of TLMs with LMPN. Some numerical examples are provided which demonstrate that the characteristic times for the echoes of the field are independent of the initial photon distribution (which is generally not discussed in the literature). These examples do not show the existence of mini echoes which are seen when this problem is treated from the viewpoint of echoes in the density matrix of the TLMs. The second subsection (Characteristic times for large numbers of TLMs and SMPN), demonstrates the characteristic times found for the photon echoes when the number of TLMs initially all prepared in the ground state (big negative spin) is much larger than the mean number of photons. In addition, it is shown that this result is independent of the initial photon distribution, which is demonstrated by a photon distribution that is initially thermal. This is contrary to what is found with small numbers of TLMs and LMPN. Section 4 continues the examination of characteristic times examining the echo behavior to greater times. The results found were unexpected but of considerable interest in that complete sets of echoes are found well past the time the initial set of echoes decays. Section 5 is an examination of the entropy of the field for this model. The calculations are only carried out to the time of the second revival, not only due to the computation time limitations but also since the behavior seen is sufficient to discuss the results. Section 6 contains the results found for the expectation value of the density matrix of the field (Q function) taken with the coherent field when the initial field is also coherent. This function has been discussed extensity in the literature, and the results seen here are both familiar and different that seen previously. A brief summary is provided.

For the interested reader, a review of the solution to the Tavis-Cummings Model Hamiltonian for the non-resonant case is provided in Appendix A. The general expressions for the time development of the ensemble averages of $E^-E^+$ is presented in an Appendix B for N two level molecules (TLMs) interacting with a single mode electromagnetic field. The TLMs are assumed to be at equivalent mode positions in the field. The general results are then simplified for all TLMs in various configurations although most of the numerical results presented in this document are for all the TLMs initially in the down state at time 0 (stimulated absorption). Exact solutions for small numbers of TLMs are provided in Appendix C for the TLMs initially in the up state. A review of the approximate solutions development for stimulated emission is found in Appendix D although the revival characteristics of this model to not completely match the exact results unless modifications are made. Approximate solutions for stimulated absorption are found in Appendix E. A reconsideration (Appendix F) of the approximate solutions found in Appendix D and Appendix E is made to obtain more exact results. The formulation for entropy for both stimulated emission and absorption can be found in Appendix G. Finally development of the results for the Q function for both stimulated emission and absorption for the initial coherent state is provided in Appendix H.

## 2 Background

Jaynes and Cummings [1] introduced the rotating wave approximation (RWA) model of a two-level atom interacting with a quantized mode of an optical cavity, with or without the presence of light (in the form of a bath of electromagnetic radiation that can cause spontaneous (stimulated) emission and absorption (the Jaynes–Cummings Model (JCM)). The JCM is and has been of great interest in atomic physics, quantum optics, and solid-state quantum information circuits both experimentally and theoretically. This was evidenced by the Tenth Rochester Conference on Coherence and Quantum Optics in 2013 in which the plenary session on the first day was devoted to the JCM and the special issue of the Journal of Physics B titled "Fifty years of Jaynes–Cummings physics" [2] where it was noted that there are nearly 15,000 articles involving Jaynes-Cummings physics. In fact the "JCM has evolved into a cornerstone of quantum state engineering" [2] especially to quantum computing and qubit theory.

Tavis and Cummings ([3-5]) extended the JCM model to any number of TLMs interacting with the electromagnetic field within the single-mode cavity. They provided exact as well as approximated solutions to the Hamiltonian. This has become known as the Tavis–Cummings model (TCM). This model has also been cited extensively although not nearly as often as the JCM but "has gained renewed interest as it can be used to implement quantum information protocols with the oscillator transferring information coherently between qubits." [6].

Using the JCM, Cummings [7] provided the theory of the time development of the ensemble averages of $E^-$ and $E^-E^+$ [for stimulated emission ($S_1$) and absorption ($S_4$)] for photon number densities representing the coherent and thermal states for the resonant case but for short times. Eberly [8-10] extended the theory to the long-term development for the ensemble average of $E^-E^+$ (for stimulated emission) showing the periodic collapse and revival of oscillations in that average. In addition, Eberly and his colleagues performed significant time behavior analysis and found a set of characteristic times for the collapse and revival of oscillations. This will be reviewed later. This author did observe long time revival of photon echoes as early as 1966 but did not characterize the time behavior while extending the results of [7] to the mixed coherent and thermal photon number distributions (unpublished). An elegant empirical demonstration of the quantum collapse and revival was observed in 1987, when a single Rydberg atom in a single mode of an electromagnetic field in a superconducting cavity was investigated [11].

Besides collapse and revival phenomena, there have been a number of papers that examined the JCM in terms of entropy [12 - 15]. The authors provide discussions of the disentanglement of the initial coherent photon state and TLM state which occurs at one half the revival time. Further, at that time, it was found that the atomic state was in a "pure" state where the entropy approched zero no matter what the intial TLM state. Tessier et al. [16] have discussed entanglement sharing for the TCMs. Jarvis et al. [17] considered up to six TLMs interacting with a coherent field with a LMPN, although displayed numerical results only for up to four TLMs. Especially interesting was the preparation of the TLMs into a so-called attractor state, which may also be called a coherent or Block state rather than the condition for which all the TLMs were initially in the all-up, all-down, or the state in which the z component of total spin is 0. Employing entropy and the Q function Jarvis demonstrated entanglement and disentanglement for the field and TLMs as well as disentangelement of the individaul TLMs at specified times proportional to the revival times. At those times the system can be written as product forms for the field and TLMs. Mini and main revivals of the probability of the TLM being in the ground state were found at times inversely proportional to the number of TLMs. However, those revivals decrease rapidly in amplitude as the number of TLMs increased. We will comment on that later. There was also a series of papers that considered the coherent spin state more fully for both the JCM model and the TC model ([18], [19]). Coupling of N TLMs to a leaky cavity was considered in [20] but the time development of the system was not discussed. One of the most recent articles discusses the dynamics of entanglement and quantum discord [21] while Feng et al. [22] considered the quantum critical behavior of the driven TCM, but for neither case more than four TLMs were considered.

Although there have been numerous papers concerning the time development for the JCM and TCM model, it appears that in all cases the number of TLMs examined in detail rarely exceeded three. Further the time development did not go beyond the collapse of the first revival except for the first works by Eberly [8-10]. Thus the extensions provided by this author appear relavent. Several papers examining stimulated emission and absorption as well as spontaneous emission of radiation in a single mode for both resonance and non-resonance for various initial photon distributions [23-25] have been published as preprints at arXiv.org. In addition, a refereed journal article corresponding to [24] was published in the special issue of Jaynes–Cummings physics [26].

References [24 and 26] provide the complete theory of the interaction between n photons and N-TLMs based on the exact orthogonal eigenvectors developed in [3 and 4] for stimulated emission, stimulated absorption, and spontaneous emission. However, a discussion of rival times was not considered in those references. Exact solutions were found for up to 4 TLMs for the resonant case while for the non-resonant case only the exact solution for 2 TLMs was found. (See appendix C) In most cases only tractable solutions for all TLMs in the up state or down state are amenable to numerical solution. However, the case of m=0 (or half the TLMs up and half down) was also considered for 0 initial photons. Approximate solutions were also provided for stimulated emission for LMPN [24] and contains numeric results for various cases. Reference [23] considered stimulated emission for 11 photon densities including those with squeezing and displacement. Several numeric examples were provided for each photon distribution. Reference [25], the progenitor of this note, considered stimulated emission and absorption.

To recap: The general solution for $E^-E^+$ for N TLMs interacting with the photon distribution as a function of time is defined as the trace of the photon number with the density matrix of the field (the TLM reduced density matrix) (see U. Fano [27]) with the results expressed in terms of the eigenstates of the TCM (Appendix B). This was first discussed in [3] Appendix F but some errors existed there. Approximate solutions for stimulated absorption can be developed even for the non-resonant case using the techniques in [3 and 4]. (Appendix E) A review of characteristic times appropriate for both stimulated emission and absorption are presented next. It will be seen that there are distinct differences for the results.

## 3 A discussion of characteristic times

### Characteristic times for few TLMs and LMPN

This section is a review of characteristic times seen for photon echoes for the case of LMPN and few numbers of TLMs. This is provided as a comparison to the results in the next section. Eberly [8-10] defined a set of characteristic times associated with the collapse and revival of oscillation for the case on 1 TLM interacting with a coherent state for a LMPN for both the resonant case and non-resonant case when the TLM was prepared in the excited state. He also provided an envelope expression for the slow decrease in the revival amplitudes. He stated that the first collapse is characterized by a Gaussian envelope with a characteristic time proportional to the inverse coupling constant between the photon annihilation and creation operators and the creation and destruction operators for the spin states (or TLMs). For the non-resonant case this envelope (in Eberly's notation) is proportional to

$$e^{\left(-\frac{2\bar{n}\lambda^2}{\Delta^2+4\bar{n}\lambda^2}\right)(\lambda t)^2}, \qquad (1)$$

where he defined $\lambda$ as the atom field coupling constant, $\Delta$ as the difference in the TLM energy and the field frequency and $\bar{n}$ is the mean photon number. To be somewhat consistent with notation used later in the paper, replace $\lambda$ by $\gamma$, define $\kappa = \gamma/\Omega$, where $\Omega$ is equivalent to $\omega_o$ (the energy separation of the TLM) used by

Eberly, the relative tuning parameter is defined as $\beta = (\omega - \Omega)/|\kappa|\Omega$, and a new definition for $\Delta$, namely that of $\Delta = \beta^2/4$, is used. Thus Eq. (1) in this notation yields

$$e^{-\frac{1}{2}\left(\frac{\bar{n}}{\Delta+\bar{n}}\right)(\gamma t)^2}$$

In this notation the characteristic collapse time is given by

$$\tau_c = \sqrt{2\frac{\Delta + \bar{n}}{\bar{n}\gamma^2}}. \qquad (2)$$

The revival times were found to be proportional to the square root of the mean photon number (for the initial photon distribution being the coherent state). This is now expressed as

$$\tau_R = 2\pi \frac{\sqrt{\bar{n} + 1 + \Delta}}{\gamma}. \qquad (3)$$

Eberly also provided an expression for a simple characteristic function expressed in terms of this paper's notation as

$$B(t) = \left[1 + \frac{(\gamma\bar{n}t)^2}{4(\Delta + \bar{n})^3}\right]^{-\frac{1}{4}}. \qquad (4)$$

Aravind and Hirschfelder [28] extensively analyzed the JCM model for the coherent resonant case and defined additional characteristic times. I reproduce the table they provided:

Table 1: Principal time scales appearing in the coherent-state JCM at resonance

| Symbol | Formula | Description |
|---|---|---|
| $t_0$ | $2\pi\omega^{-1}$ | Period of external field |
| $t_R$ | $\pi(\sqrt{\bar{n}}\gamma)^{-1}$ | Period of the Rabi Oscillation |
| $t_c^f = \tau_c$ [1] | $(2\gamma)^{-1}$ | Collapse time of inversion (fast coherence) |
| $t_E^f = \tau_R$ | $4\pi\sqrt{\bar{n}}t_c^f = 2\pi\sqrt{\bar{n}}(\gamma)^{-1}$ | Echo time of inversion (fast coherence) |
| $2t_E^f$ | $8\pi\sqrt{\bar{n}}t_c^f = 4\pi\sqrt{\bar{n}}(\gamma)^{-1}$ | Oscillation period of slow coherence |
| $t_c^s$ | $2\bar{n}(\gamma)^{-1}$ | Collapse time of slow coherence |
| $t_q$ | $\pi\bar{n}(\gamma)^{-1}$ | Boundary of quasi-periodic regime |
| $t_E^s$ | $4\pi\sqrt{\bar{n}}t_c^s = 8\pi n\sqrt{\bar{n}}(\gamma)^{-1}$ | Echo time of slow coherence |

Four of the times, $t_0, t_R, \tau_c,$ and $\tau_R$ are easily recognized from the work done by Eberly as the characteristic times associated with the collapse and revival of the ensemble average of $E^- E^+$ or the inversion of the TLM.

---

[1] Note that the characteristic time provide by Eberly is different than that provided in the table. The collapse time given by Eberly is $\frac{\sqrt{2}}{\gamma}$.

The fifth time, $t_q$, designates the end of the coherent collapse and revivals and the beginning of the chaotic behavior. The three other times listed are not related to the same behavior in the inversion of the TLM. Instead it is noted that Arivind [28] considered the spin vector of the TLM converted to a rotating coordinate system such that at resonance, the spin component $S_1'$ vanished, the inversion component remained identical to the inversion in the non-rotating frame and the $S_2'$ component rotates with both a slow and fast component. The three remaining times in the table are all related to the slow oscillation in $S_2'$ with $2t_E^f$ being the oscillation period of the slow oscillation, $t_c^s$ being the collapse time of the slow oscillation, and $t_E^s$ the echo time for the slow oscillation. This author will not discuss these spin components nor their corresponding characteristic times further in this paper.

This completes the review of the classic revival times; however, before proceeding to the discussion of the revival times associated with big negative spin and SMPN, there are two topics of interest that deserve comment. The first concerns the use of the initial photon distribution used in various papers concerning echo behavior and the second is a brief observation about the existence of mini revivals.

To my knowledge, there has not been any discussion of characteristic times for the case of a single TLM coupled with a non-coherent photon distributions. In the literature, all cases discussed assumed the use of the coherent photon density matrix. It turns out that that assumption is not always needed for the correctness of (3). This is demonstrated with the following examples provided in Table 2. Note that the photon density distribution as well as the time development for $S_1$ is shown. $S_1$ represents the time dependent part of $E^-E^+$ in the Cummings notation [7] and is given for 1 TLM initially in the up state (Eq. (C1)) as

$$S_1\left(\bar{n}, 1, \frac{t}{\tau_R}, \Delta\right)_{NR} = \sum_{n=0}^{\infty} \langle n|\rho_f(0)|n\rangle \left(\frac{n+1}{n+1+\Delta}\right) Sin^2\left[2\pi\sqrt{\bar{n}+1+\Delta}\sqrt{n+1+\Delta}\frac{t}{\tau_R}\right].$$

The subscript on the S means that the TLMs are initially in the up state. The subscript NR means non-resonant. If $\Delta$ is 0, then the results are for the resonant case. The density matrix of the field at time 0 is given by $\langle n|\rho_f(0)|n\rangle$ and is provided by the left-hand column in Table 2 and described in more detail in [23]. This expression is obtained by expressing $E^-E^+$ in terms of the trace of the photon number with the TLM reduced density matrix of the field at time t [27]. This density matrix is found by taking the unitary transformation of the field density matrix at time 0 with the unitary operator being the exponential of the TCM Hamiltonian. The resulting general expression is then expanded using the identity matrix in terms of the eigenvectors of the TCM model. The process is straightforward but tedious [3, 24, 26] and Appendix B below. The general expression is difficult to evaluate numerically except for special cases such as setting all the TLM initially in the up state, the down state, or with half the TLMs in the up state and half in the down state.

The top row for Table 2 is for the thermal distribution, shown on the left, with 24 mean thermal photons. The corresponding $S_1$ shows photon echoes assuming non-resonance with $\Delta = 100$. The second row is for the squeezed coherent initial photon distribution with a squeezing parameter r=2, the number of coherent photons of 899, and a mean photon number of 912. The echo behavior is for the resonant case. The third row contains the results for the squeezed Fock state with a squeezing parameter r=1 and the initial photon number of 20, with a resulting mean photon number of 76. The non-resonance parameter $\Delta = 148$. The forth row is for the displaced squeezed thermal state with an initial number of thermal photons, a squeezing parameter of 1 and a displacement of 1000 coherent photons with the result of 1039 mean photon number. The fifth row is for the displaced number state with initial photon number is 25 and a displacement equivalent to 1000 coherent photons. This results in a mean photon number of 1025. The last row is for the squeezed displaced number state, with an initial photon number of 25, a displacement of 1000 equivalent coherent photons, and squeezing parameter r=1. The resulting mean photon number is 1095. The last three rows are for resonance. For more details concerning the initial photon densities see [23] as stated above. All calculations for $S_1$ are carried out to 3 standard deviations in n for the photon distribution.

It is immediately seen that for all examples given, the revival time expressed in Eq. (3) is correct except for row 3 of the table below. The echo time for that row is $\tau_R/2$; however, for that case the photon density only has valid values for even photon number. It should also be noted that the echoes are not necessarily symmetric and the echo times refer to the center of the echo and not the peak. Further, the behavior of the collapse also appears to be dependent on the shape of the photon density.

It has also been found that the main revival times for TLM numbers greater than 1 are also given by Eq. (3) as long as the number of photons is much greater than the number of TLMs. In fact a good approximation for the time dependent part of $E^-E^+$ for higher TLM number is just the value of $S_1$ for a single TLM multiplied by the number of TLMs.

*Table 2: Demonstration that the Eberly Echo time is correct for other initial photon density matrices for most cases.*

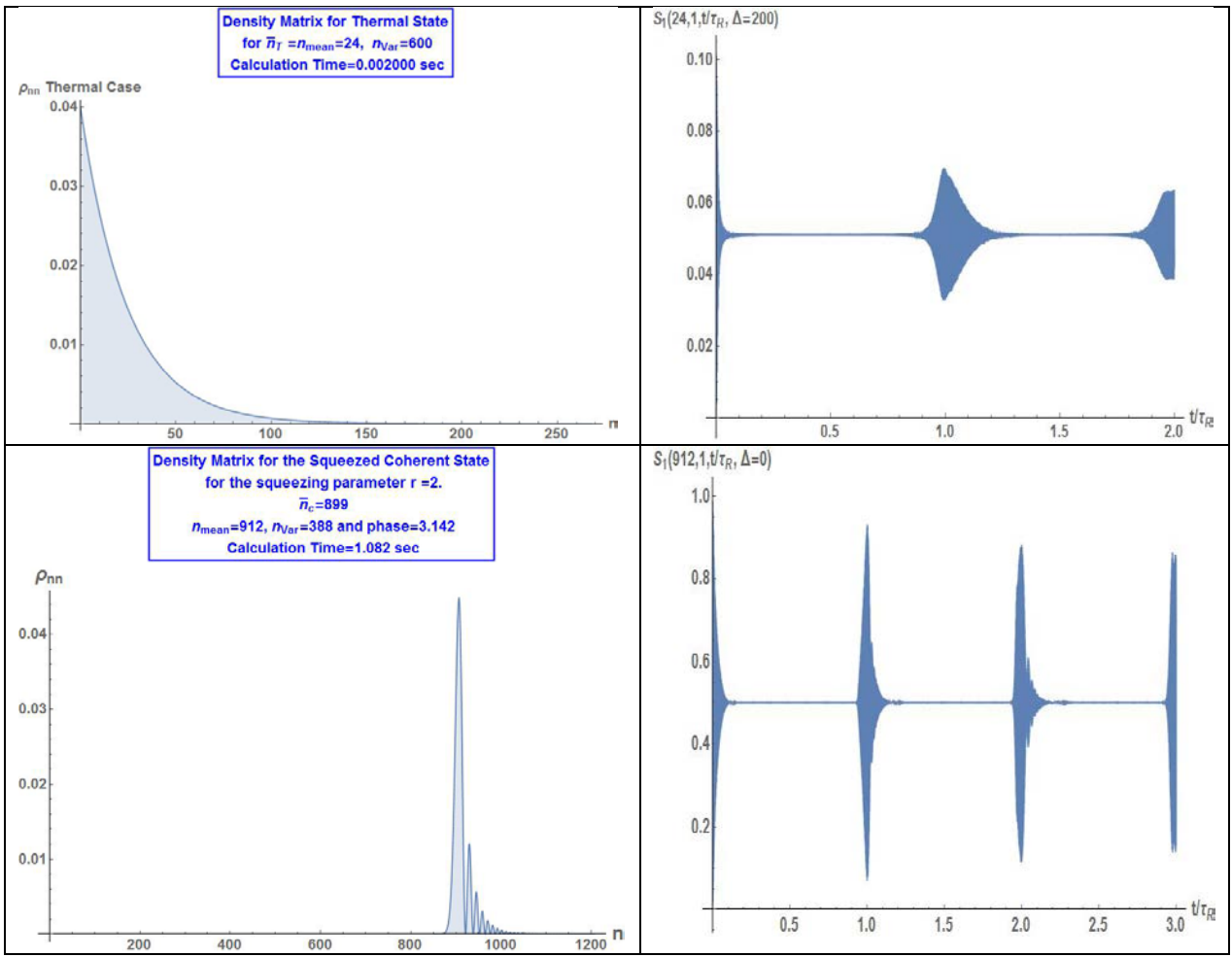

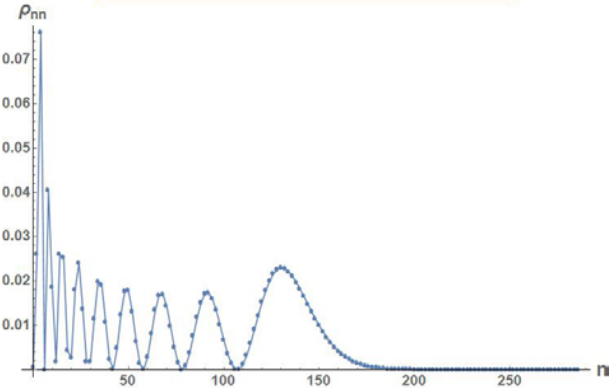
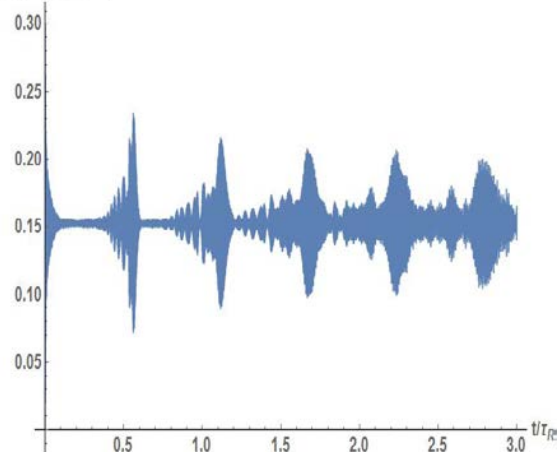
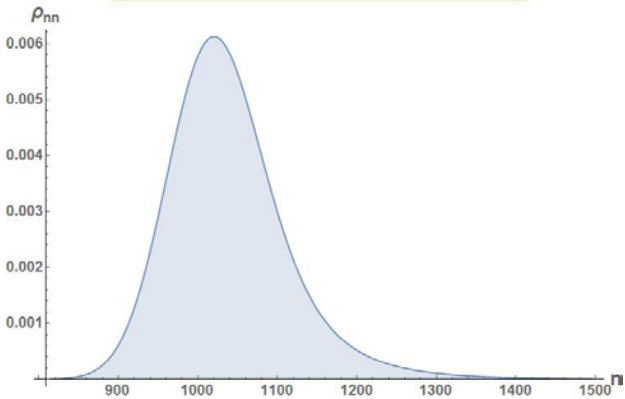
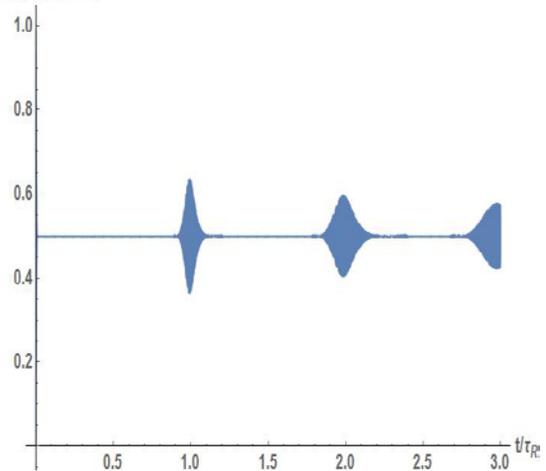
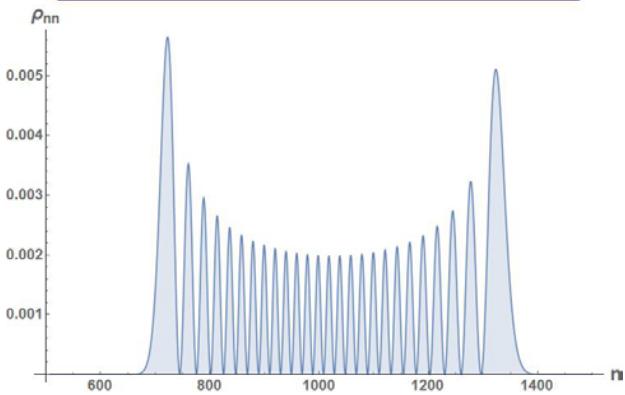
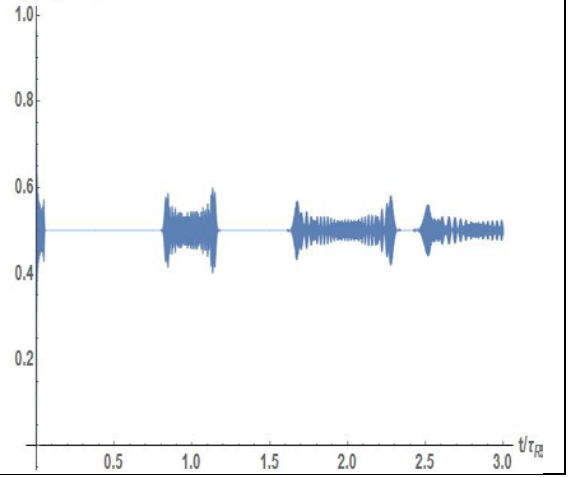

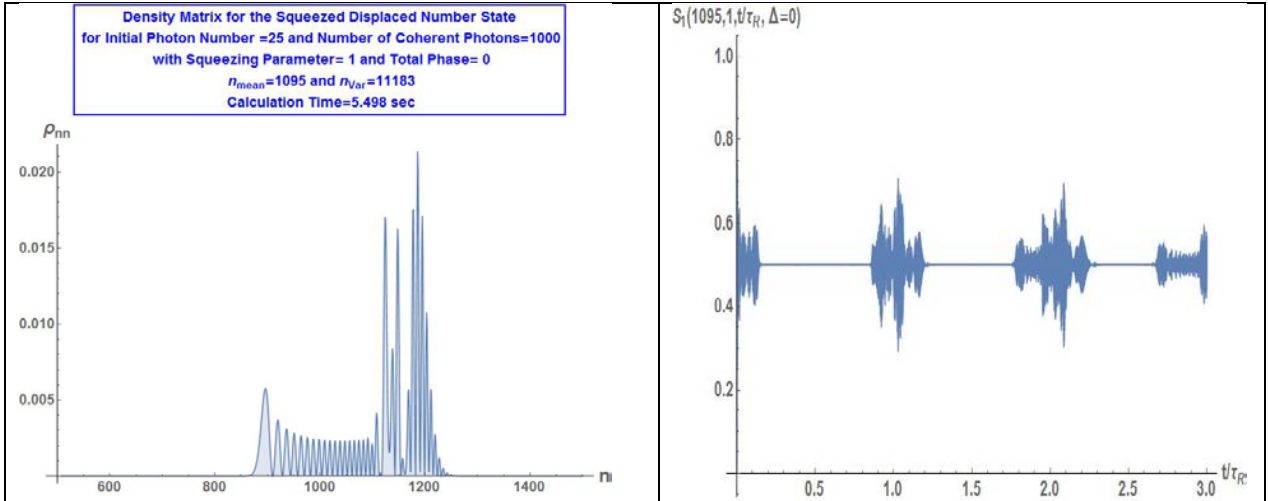

Before moving on to the second subject of interest, the equation for $S_1$ can be used to examine the other characteristic times. Table 3 is used for that purpose. All calculations were performed for a single TLM and a mean photon number of 899. The bottom right hand example was for a resonance parameter specified by $\Delta = 100$.

*Table 3: Demonstration that the other characteristic times defined by Eberly are correct*

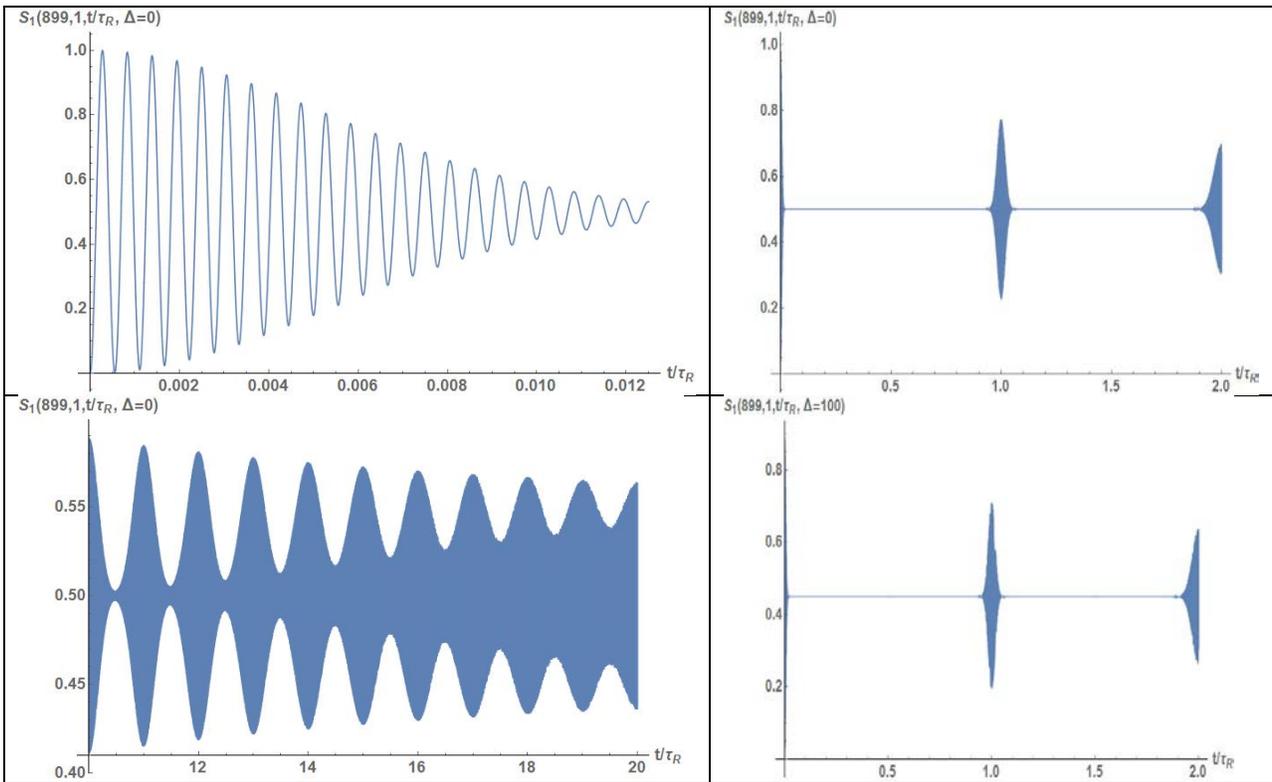

As can be seen for both the resonant and non-resonant cases (right hand side) the echoes appear at integer values of $\tau_R$. In addition the collapse time seen in Eq.(2) or the third row of Table 1 agrees with the number of oscillations (number of Rabi periods) shown in the upper left corner of the table. By that time the amplitude has collapsed to a value of $e^{-1}$. In the lower left corner, the number of echoes reached before significant merging has reached 13 also agreeing with $t_q$ given in Table 1.

The second subject of special interest in this section is the existence of mini revivals or echoes. It is found that mini revivals are not seen for $E^-E^+$. On the other hand Jarvis [29] and Everitt [30] and references therein have suggested that the revival times are inversely proportional to the number of TLMs. In those papers, it was assumed that there was a LMPN and that the initial photon state was coherent. The discussions in those papers focused on entanglement and attractor states. Their findings suggest that the initial state of the TLMs play a significant role in echo timing and the onset of disentanglement. There was little discussion of the cases when the TLMs were completely in the up or down states. Their echo analysis focused on the field-reduced density matrix and the probability that the TLMs were in the ground state as a function of time. It was indeed found that mini revivals were seen at integer values of $\tau_R/N$ but the amplitude of those mini and main revivals decreased as N increased. Variations of the methods referenced in this paper [24 and 26] can be used to verify those results. It should be pointed out that at least one reference considered the case of a small number of TLMs with a SMPN and solved for both the exact solution and mesoscopic field approximation [31].

In the following table, exact results for $S_1$ are shown for the photon echoes rather than TLM echoes for a single TLM, 2 TLMs, and 4 TLMs all initially in the up state and all calculated with an initial coherent distribution with a mean photon number of 899. The results are plotted using normalized time (Eq. 3) for the resonant case except for 1TLM where an extra figure is included for non-resonance with $\Delta = 100$. As can be seen there exists no mini revivals on the amplitude scale for the main photon revivals on the scale seen for the main revivals.

*Table 4: Demonstration that mini photon echoes are not seen*

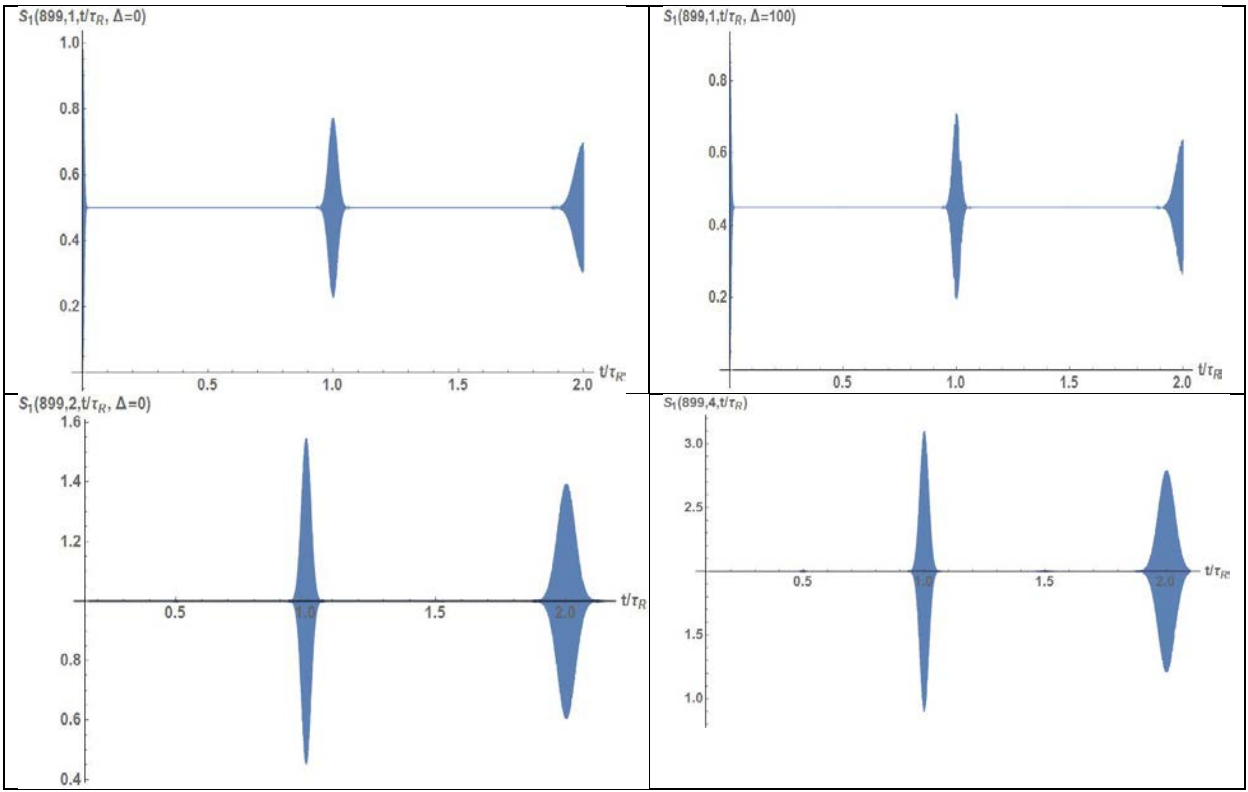

As can be seen for both the resonant and non-resonant cases (right hand side) the echoes appear at integer values of $\tau_R$

To continue the examination of characteristic times, we next consider the case of 2 TLMs with reference to [29] interacting with the coherent density matrix. Again we assume that the initial photon distribution is coherent with a LMPN. We first consider both TLMs in the down state and secondly consider the case of an equal admixture to 2 TLMs in the up state with 2 TLMs in the down state. We use the exact solutions shown in Appendix C, namely (C2) and (C6). Recall that when the TLMs are in the up state energy is added to the photon field while when the initial states are down they subtract. What is found (See Table 5) is that there is indeed a mini revival at $\tau_R/2$ but it is very small and would not have been seen on the scale of the main revivals. The results for the difference is also not of particular interest and the results is small and shows only the main revival times

*Table 5: Examination of revival times for 2 TLMs*

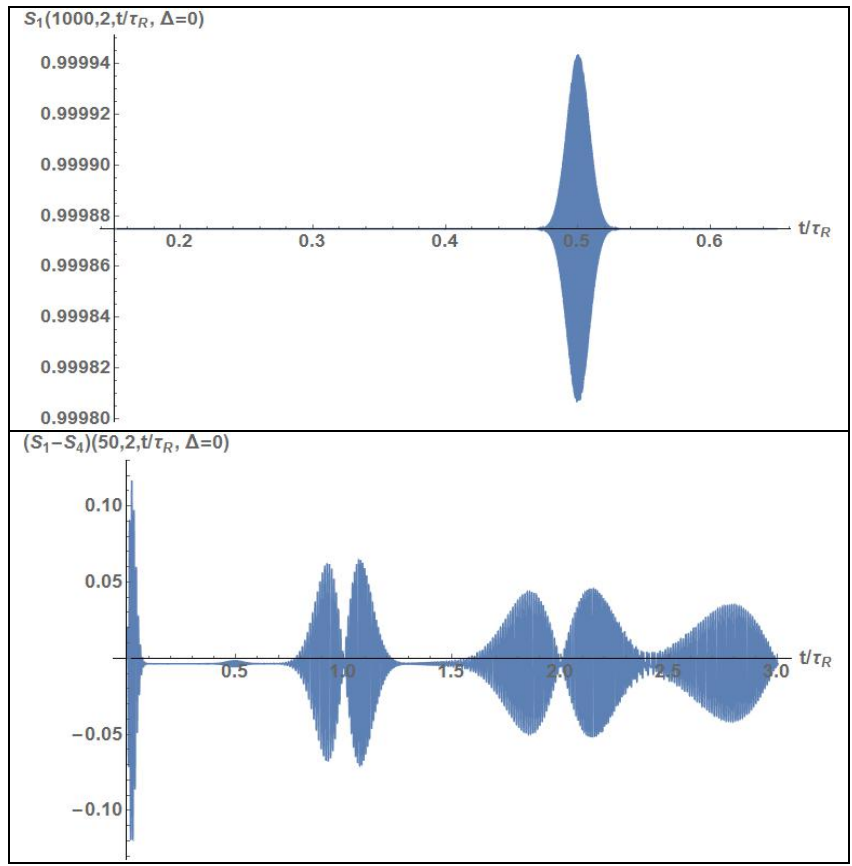

We have indeed carried out the analysis to larger numbers of TLMs. As in the case of 2 TLMs, there are some mini revival times which are barely perceptible and not of the same interest as those found for the examination of revival times found when the analysis is carried out for the TLM density matrix such as the probability that the TLMs are in the ground state when they were initially prepared in that state. Thus for photon echoes with a LMPN ≫ than the number of TLMs the results mimic the single TLM case. This completes the review of characteristic times and the 2 items of special interest. In the next section, we return to the consideration of revival times but for large numbers of TLMs all in the down state and an initial photon density with a SMPN.

## Characteristic times for large numbers of TLMs and SMPN

This section again returns to the consideration of the characteristic times but this time for the new results found for a large number of TLMs all in the ground state (big negative spin) interacting with an initial photon density with few photons. Again we are interested in the temporal behavior of $E^- E^+$. It is much more difficult to obtain definitive algebraic results for the characteristic times for this case and we instead rely on numerical results for clues about the behavior. For that reason the following section addresses the numerical results and we come back to the question of characteristic times as the numerical results are presented.

In order to examine the characteristic times for the big negative spin with few mean photons requires some detailed numerical calculations to verify assumptions. Toward that end various cases were selected for study. The calculations are time consuming thus only a few numerical results were considered. This author has provided numerous examples for the various photon densities in [23] for stimulated emission and some examples for stimulated absorption in [25]. Unfortunately those results were not very useful for this study. From the development of the approximate results see Appendix D and Appendix F. The approximate results were developed with the hope of obtaining equations with sufficient accuracy and simplicity that calculations could be performed in shorter times. During the development of those approximate expressions, it became apparent that the characteristic revival time could be represented, using the example by Eberly [8] as

$$\Omega|\kappa|t = 4\pi\sqrt{N+\Delta}\tau.$$

With this we rewrite the exact and approximate solutions of $S_4$ (B13 and E20) as

$$S_4(\bar{n}, N, \gamma t) = -4 \sum_{n=0}^{\infty} \langle n|\rho_f(0)|n\rangle \sum_{j=0}^{Min[N,n]-1} \sum_{j'=j+1}^{Min[N,n]} Sin^2\left\{4\pi\sqrt{N+\Delta}\frac{\left[q_{\frac{N}{2},(n-\frac{N}{2}),j} - q_{\frac{N}{2},(n-\frac{N}{2}),j'}\right]}{2}\tau\right\}$$
$$\times (A^*)_n^{\frac{N}{2},n-\frac{N}{2},j} A_n^{\frac{N}{2},n-\frac{N}{2},j'} \sum_{p=0}^{min[N,n]} pA_{n-p}^{\frac{N}{2},n-\frac{N}{2},j}(A^*)_{n-p}^{\frac{N}{2},n-\frac{N}{2},j'} \text{ and}$$

(5)

$$S_4(\widetilde{\bar{n},N},\gamma t) \cong \sum_{n=0}^{\infty} \langle n|\rho_f(0)|n\rangle n \frac{N-\frac{n}{2}}{N-\frac{n}{2}+\Delta} Sin^2\left\{4\pi\sqrt{N+\Delta}\sqrt{N-\frac{n}{2}+\Delta}\tau\right\}. \quad (6)$$

Note that in Eqs. (5 and 6) above, N is the number of TLMs, $\bar{n}$ is the mean number of photons in the initial photon distribution, $q$ is the adjusted eigenvalue and the $As$ are the coefficients of the eigenvectors of the TCM Hamiltonian. One should also note that the subscript 4 on S represents the fact that all the TLMs are initially in the down state. This notation was also used by Cummings [7] when he considered the time development of a single TLM in the down state interacting with the initial photon field.

In order to test the hypothesis for the revival time, assume the resonant case ($\Delta= 0$), N=900, the coherent photon density is

$$\langle n|\rho_f(0)|n\rangle = \rho_{nn} = \frac{e^{-\beta^2}|\beta|^{2n}}{n!}, \bar{n} = |\beta|^2.$$

Choose $\bar{n} = 10$ to insure that the number of TLMs >>n. The exact and approximate solutions within the numerical accuracy achieved by limiting the upper limit of the summation over n to 35 is shown in Figure 1.

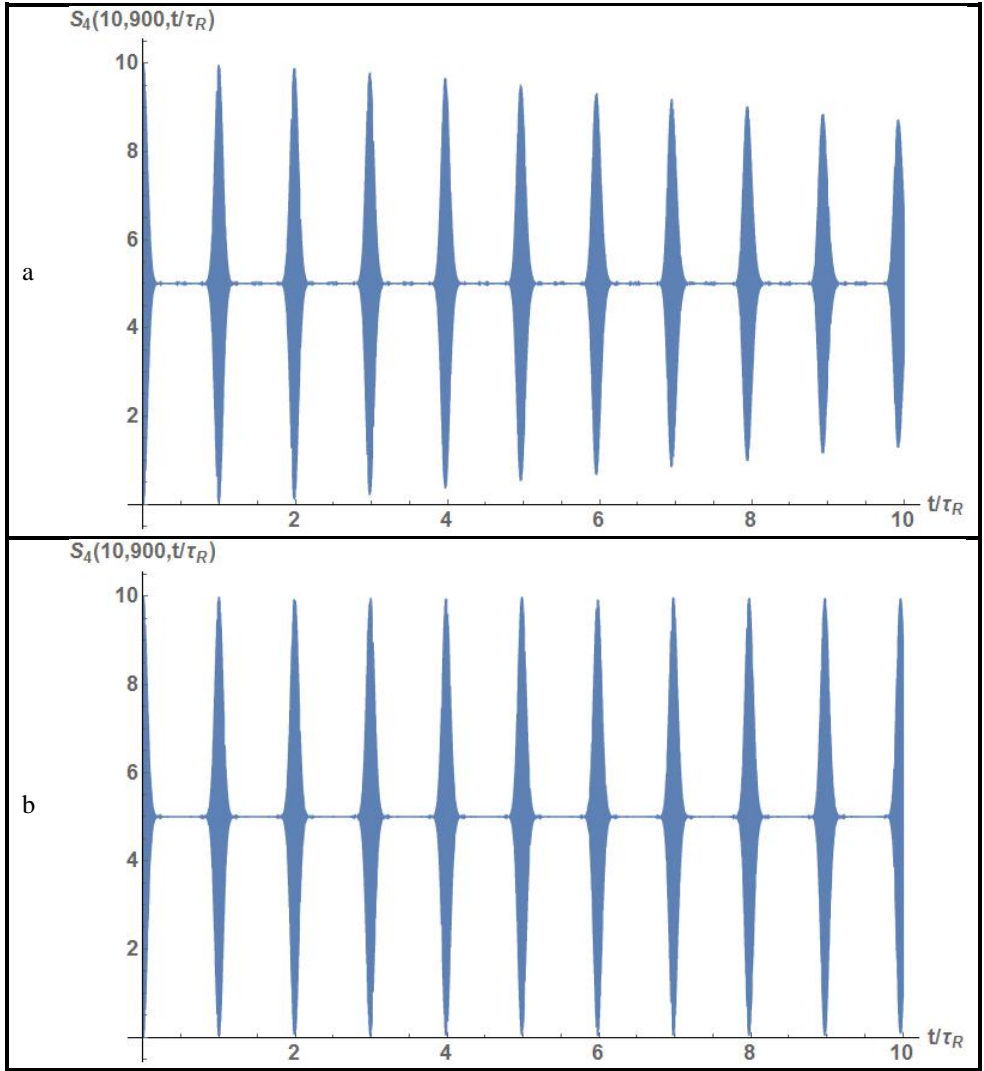

Fig. 1: Exact (a) and approximate (b) results for stimulated absorption for 900 TLMs and a mean photon number of 10 for the coherent photon distribution.

As a second example, choose the thermal photon density represented by

$$\langle n|\rho_f(0)|n\rangle = \rho_{nn} = \frac{1}{\bar{n}_T + 1}\left(\frac{\bar{n}_T}{\bar{n}_T + 1}\right)^n, \bar{n} = \bar{n}_T.$$

Note that for this case the variance in the number of photons is considerably greater than for the coherent case, namely $var = \bar{n}^2{}_T + \bar{n}_T$ rather than $|\beta|^2$. That is $var$ =110 rather than 10 which means that the summation over n in Eqs. [5](#) has to be carried out to larger values. Again choose $\Delta$= 0, N=900 and $\bar{n} = \bar{n}_T = 10$. The exact and approximate results for these calculations are presented in Fig. 2.

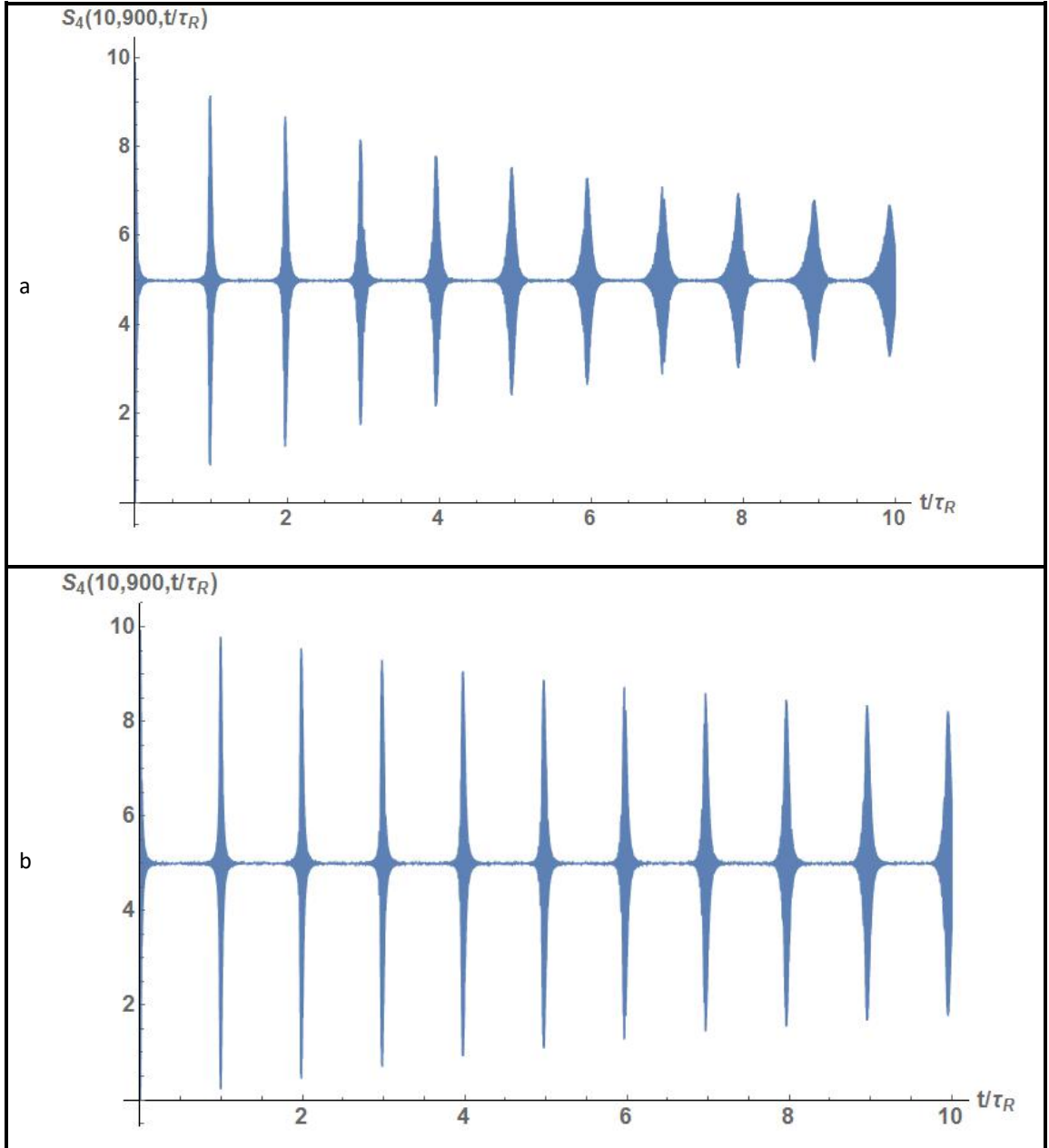

*Fig. 2: Exact (a) and approximate (b) results for stimulated absorption for 900 TLMs and a mean photon number of 10 for the Thermal photon distribution.*

From the two examples given there is agreement that the revival times are indeed $4\pi\sqrt{N+\Delta}/\gamma$ and it is seen that the time behavior between the exact and approximate solutions are nearly identical (except that for the thermal distribution, the width of the revivals for the exact case becomes larger for larger time). On the other hand the amplitudes of the revivals for the approximate solutions do not fall as quickly as for the exact solution. Please note that the results shown in Fig. 2 is surprising in that it is completely contrary to calculations for stimulated emission for the thermal distribution. We are seeing revivals while for stimulated emission the results would be chaotic. We have seen that the revival times agree with the revival time as discussed above. Other characteristic times are of interest, namely the equivalent Rabi period, the initial collapse time, and the time over which the revivals decay into the background.

The Rabi period for large numbers of TLMs all initially in the down state interacting with a SMPN is found by carefully examining the near 0 time oscillations and measuring the period from 0 out to the second zero in the oscillations. It is found that

$$t_R = \frac{2\pi}{\Omega|\kappa|\sqrt{N}}..  \quad (7)$$

This should be compared to the results in Table 1. It is also possible to estimate what the collapse time of the initial peak or widths of subsequent revivals.

Toward that end, calculate the time behavior for both the coherent and thermal distributions out to a normalized time of 1.5 for 900 TLMs and a mean photon number of 10.

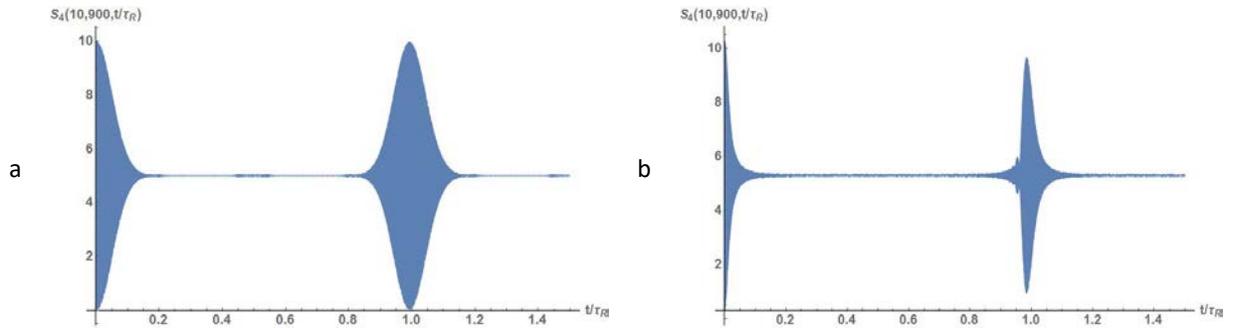

Fig. 3: Exact solution for 900 TLMs for a mean photon number of 10 for both the coherent (a) and thermal (b) photon densities.

We have closely examined the coherent results for various numbers of TLMs and mean photon number. Interestingly the results always look like the left case in Figure 3, indicating that the collapse and revival characteristics seems to be nearly independent of those quantities as long as the number of TLMs is always greater than the mean photon number when plotted using normalized time. For instance, the collapse time and revival shapes appear to follow an exponential with time dependence varying between $1/10$ and $1/4\pi$ for N between 169 and 1600; for example, $\exp(-\left(4\pi \frac{t}{\tau_R}\right)^2)$. It is also apparent that the shape of the collapse and the shape of the revivals is dependent on the initial photon distribution. One other observation is that the revival time $4\pi\sqrt{N+\Delta}/\Omega|\kappa| = 4\pi\sqrt{N+\Delta}/\gamma$ is not exact if N gets too small compared to the mean number of photons. Examination suggests that a more accurate results may be

$$\tau_R = \frac{4\pi\sqrt{N - \frac{\bar{n}}{2} + \Delta}}{\Omega|\kappa|}. \quad (8)$$

Finally we determine the time over which the revivals collapse. If the results in Table 1 is any indication an equivalent time may be

$$t_c^s \approx \frac{N}{\Omega|\kappa|} \text{ or in terms of } \frac{t}{\tau_r} \approx \sqrt{N}. \quad (9)$$

In order to examine this question numerically, consider 2 cases for N=169 and N=1600 for a mean photon number of 10 and the coherent case.

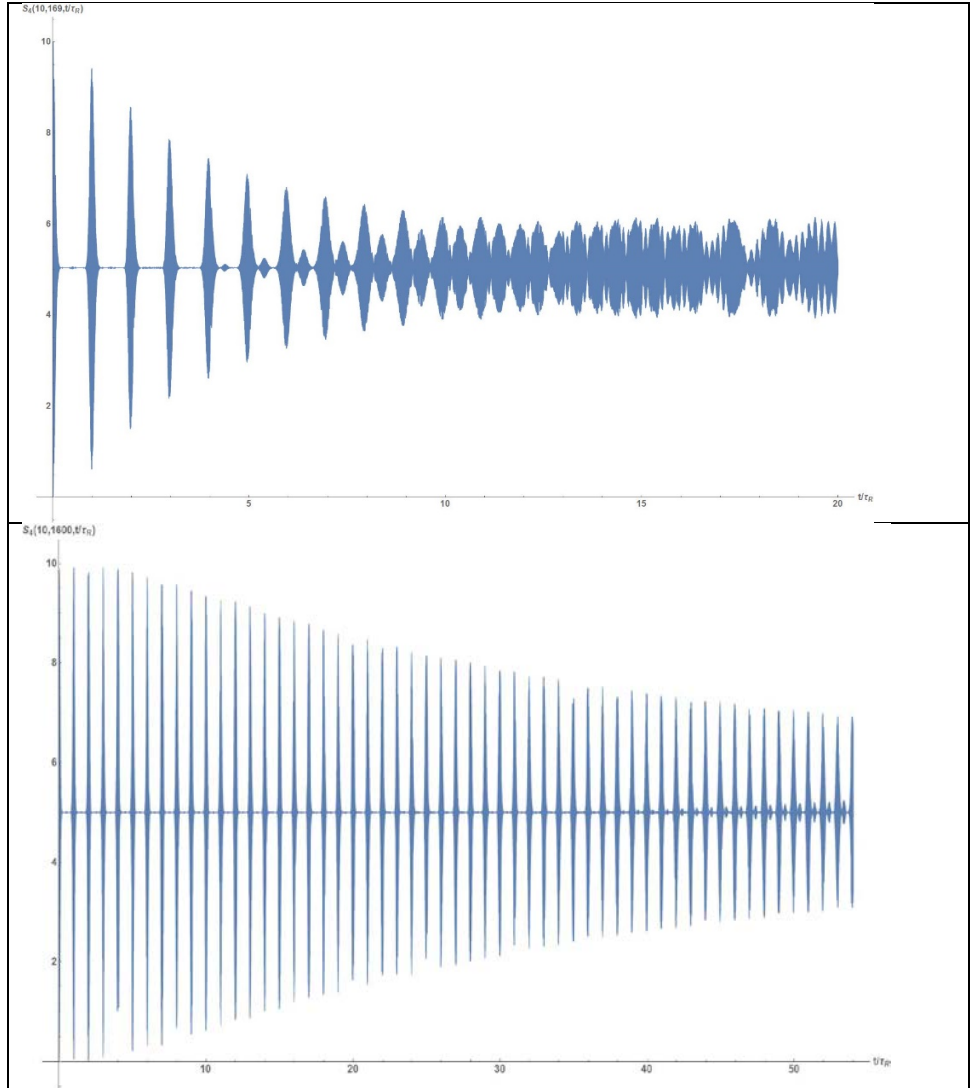

Fig. 4: Examination of the collapse of revivals as a function of $t/\tau_R$ for an initial coherent photon density with a mean photon number of 10 for 169 TLMs (top) and 1600 TLMs bottom. Note the beginnings of secondary revivals which grow and then merge into the main revivals.

For N=169 TLMs we see the beginning of secondary revivals between the 4$^{th}$ and 5$^{th}$ main revivals. By the 10$^{th}$ revival the secondary revivals have grown to rival the main revivals and by $\sqrt{169}$ = 13, these secondary revivals have merged with the main revivals and it is difficult to separate them. For N=1600, on the other hand, the secondary revivals do not become apparent until about the 40$^{th}$ revival, which is equivalent to $\sqrt{1600}$. Within another 20 revivals the secondary revivals are again merged. We also note that the secondary revivals do not appear exactly 1/2 way between the main revivals but are perhaps 1/3 to 1/4 of the way between the main revivals. In the section above the equivalent Rabi period, the collapse period and the period over which the revivals collapse into the back ground have been discussed. These results should be compared to the equivalent times shown in Table 1 for the LMPN and small numbers of TLMs.

As mentioned earlier, the exact and approximate results do not appear to provide identical results (as seen in Figures 1 and 2) as time increases. For that reason a more detailed examination reveals that Eq. (6) is insufficient to mimic the exact results and for that reason a more detailed examination of this discrepancy is provided in Appendix F.

It is difficult to physically interpret the numerical results presented in this section as evidenced by Fig. (2) and the characteristic times presented. The results appear almost identical to the case of a small number of TLMs and a LMPN in that there is a quick collapse followed by a long period where the function $S_4$ does not oscillate followed by slowly decreasing revivals with intermediate non oscillating regions. Due to the similarity to the more familiar example, one is tempted to speculate that the physical discussion in [14] also applies here. That is, that the region between the initial collapse and first revival and the intervals between revivals represents a state in which the TLMs achieve almost perfect coherence while at the revival regions they do not but instead the field and TLMs become disentangled as discussed later. It may also be speculated that the initial decay of revivals is due to the eventual dephasing on the TLMs. However, this does not explain the re-occurrence of revival sets as seen in the next section. One further point made in [14], was that any initial state of the TLMs would result in the occurrence of a pure TLM state between the initial collapse and first revival. For the model here, with a large number of TLMs and few photons, this would not be the case if the TLMs were started in the all up state.

## 4 Examination of echo behavior at greater times

A more exact approximate solution provided by Eq. (F2) for non-resonance while the simplified version for resonance for the coherent photon distribution is used to calculate the results shown below, namely:

$$S_4(\widetilde{\bar{n}, N}, \gamma t) \cong \sum_{n=0}^{\infty} \frac{n}{2^{n-1}} \langle n|\rho_f(0)|n\rangle \sum_{j=0}^{Min[N,n]-1} Sin^2\left\{2\pi\sqrt{N}\left[q_{\frac{N}{2},\left(n-\frac{N}{2}\right),j} - q_{\frac{N}{2},\left(n-\frac{N}{2}\right),j+1}\right]\tau\right\}\frac{(n-1)!}{j!(n-j-1)!}. \quad (6a)$$

The basis of the derivation for the resonant case is found in [3 and 5] and is known as the Modified TLM Approach. Basically the TCM Hamiltonian was solved again by making approximations assuming that nearly all the TLMs are in the down state and that there are few photons. It was found that for Eq. (5), only terms with $j' = j \pm 1$ were of significance. Secondly the components of the eigenvectors of the Hamiltonian could be expressed in terms of the Hypergeometric function. The mistake made in Eq. (6) was the assumption that the eigenvalues were an exact linear function of the eigenvector index j. Instead the more correct approximation Eq. (6a) used of the exact eigenvalues found for the exact solution of the TCM Hamiltonian. It is pointed out that the q's are not the eigenvalues but are directly related by $q = (c - \lambda)/|\kappa|$, where $\lambda$ is the eigenvalue, c is the good quantum number n+m, and $\kappa$ was defined above as $\gamma/\Omega$. Recall that n is the number of photons and m=$-N/2$ when all the TLMs are in the down state. To demonstrate the correspondence between the exact expression and this approximate expression, see Figure 5 below.

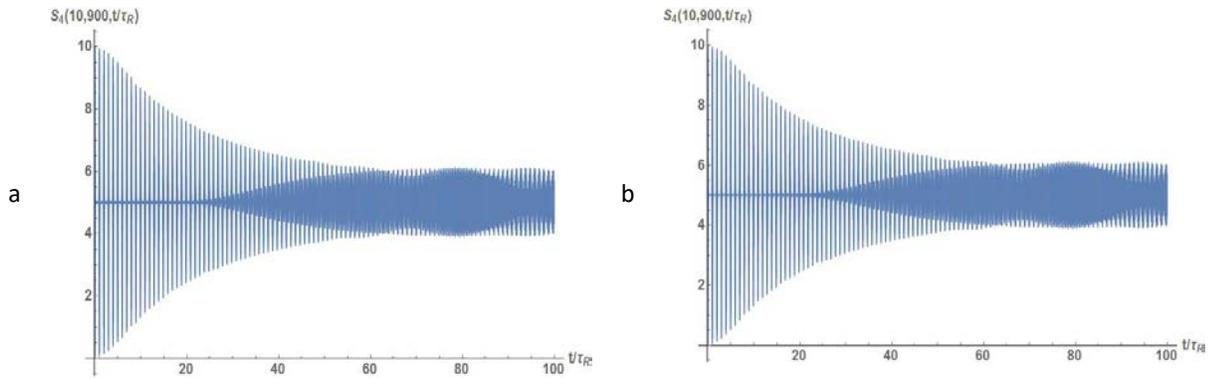

Fig. 5: Comparison between the exact expressions for $S_4(\bar{n}, N, \gamma t)$ (a) and the approximate expression using Eq. (6a) $S_4(\widetilde{\bar{n}, N}, \gamma t)$(b) for the resonant case not ignoring the variation in eigenvalues as a function of j.

The two results appears to be almost identical. This shows that the modification to the approximate results was successful, providing very good results and with a much faster computational speed. As an exercise, calculate the approximate solution out to a normalized time $\tau = 800$.

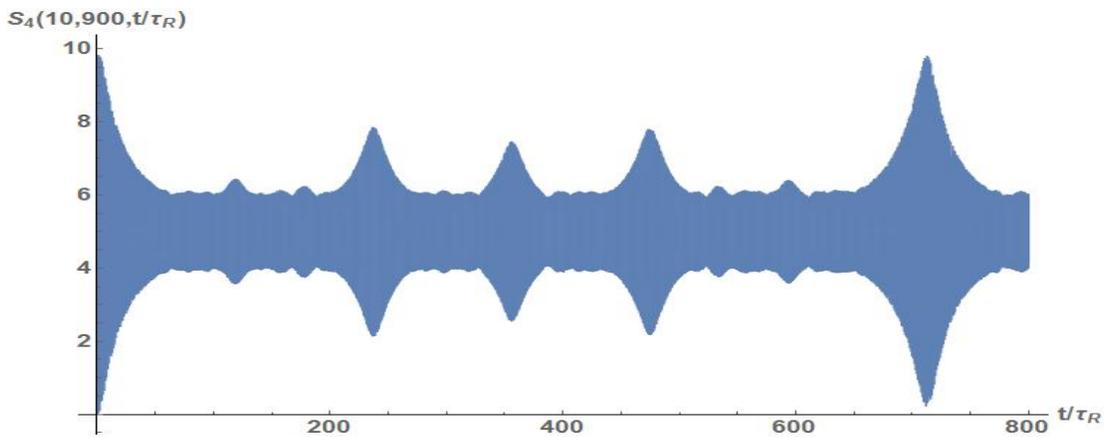

Fig. 6: Approximate results for stimulated absorption for 900 TLMs and a mean photon number of 10 for the coherent photon distribution for normalized time out to 800 assuming resonance.

This is certainly not the expected results, as it appears that there are apparent revivals asserting at larger times which are, in fact, highly symmetric. Whether or not these are primary, secondary, or other revivals is unclear. To verify that the calculated results are not an artifact of using the approximate solution, the calculation was rerun for both the exact and approximate solutions for normalized times $190 \leq \tau \leq 290$.

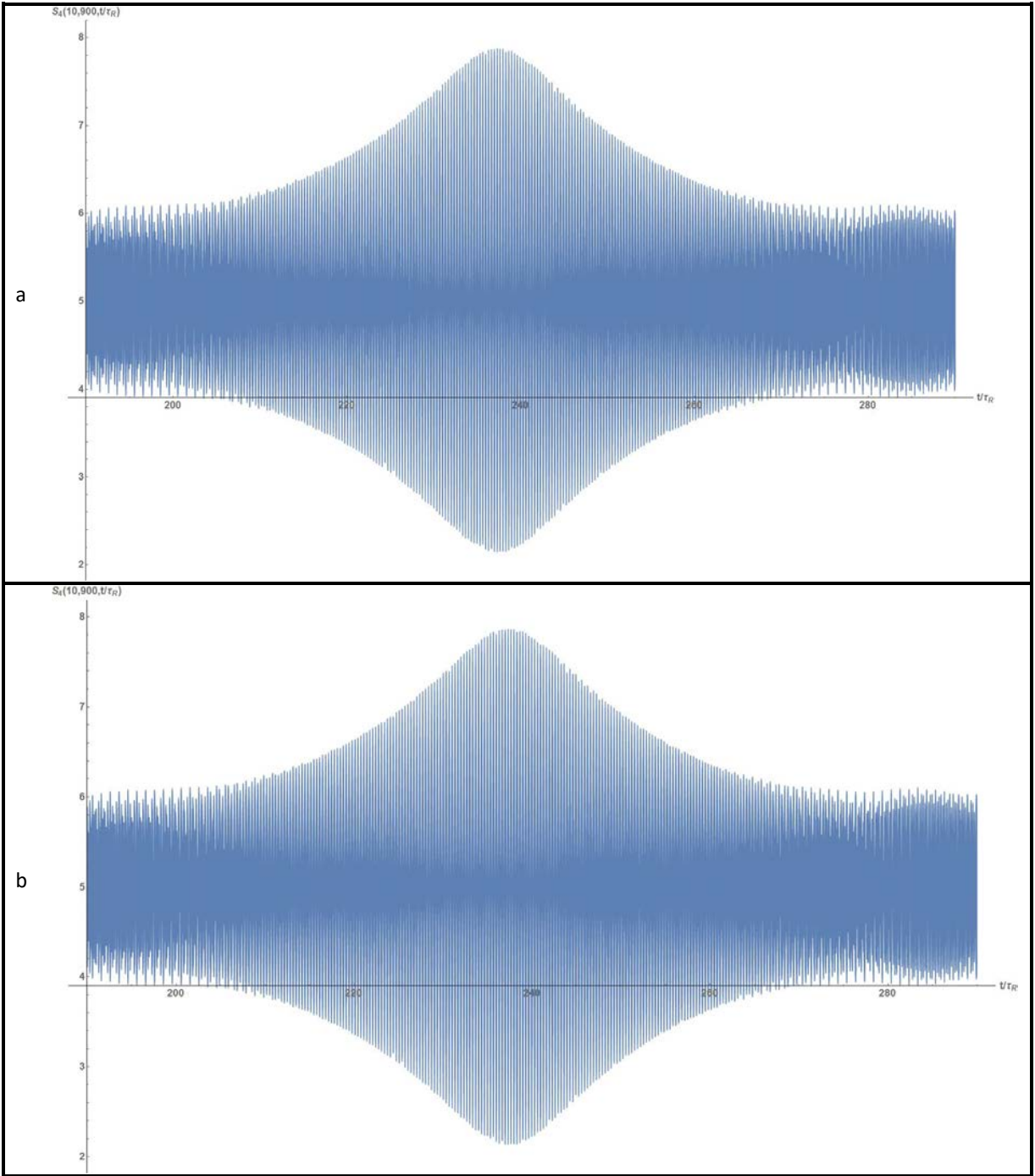

*Fig. 7: Exact (a) and approximate (b) results for stimulated absorption for 900 TLMs and a mean photon number of 10 for the coherent photon distribution for normalized time between 190 and 290 assuming resonance.*

The results indicate that the results are not an artifact of using the approximate solution. A count of revivals shows that during the time interval of 190-290 about 30 revivals is found in every 10 units of $\tau_R$ while in another interval between 340 to 370 there are 20 revivals during each 10 units of $\tau_R$; however, the revivals do not fall on integer values or 1/2 integer values of $\tau_R$. In fact a careful examination of the interval between 0 and 750 has shown some areas presenting peaks that contain as many as 50 revivals within a 10 unit interval. Thus it appears that these are not the prime revivals. To obtain more detail of the peak seen near $\tau$ =712, the exact calculation was run for resonance for normalized time between 665 and 765. The results are shown in Figure 8.

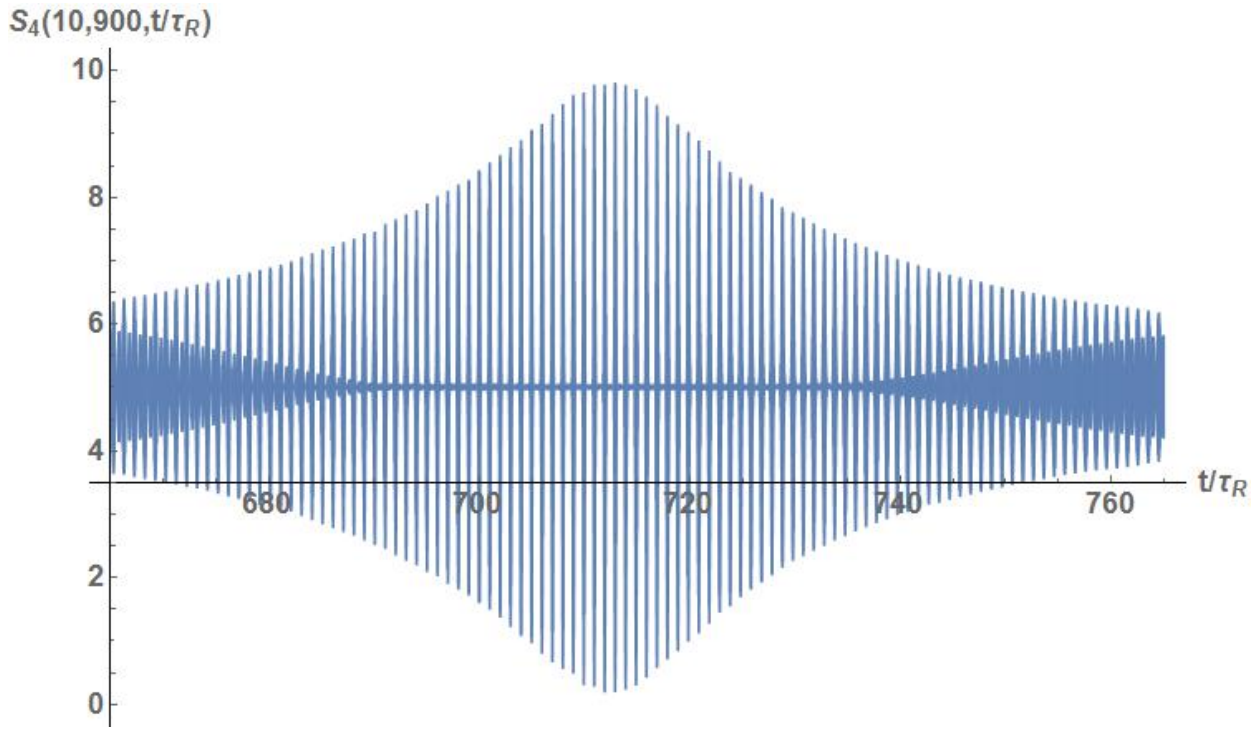

Figure 8: Exact results for stimulated absorption for 900 TLMs and a mean photon number of 10 for the coherent photon distribution for normalized time between 665 and 765 assuming resonance.

Amazingly this looks like the results for $S_4$ for time near 0 but is, in fact, symmetric about the time of approximately 712. The only visible difference between this and the approximate results is that the revivals appear to be a little thicker. Similar results have been found for the thermal photon distribution. However, for that example, the main revival did not completely recover to a maximum value of 10. To demonstrate this, the set of results for the thermal photon density distribution is shown in Figure 9

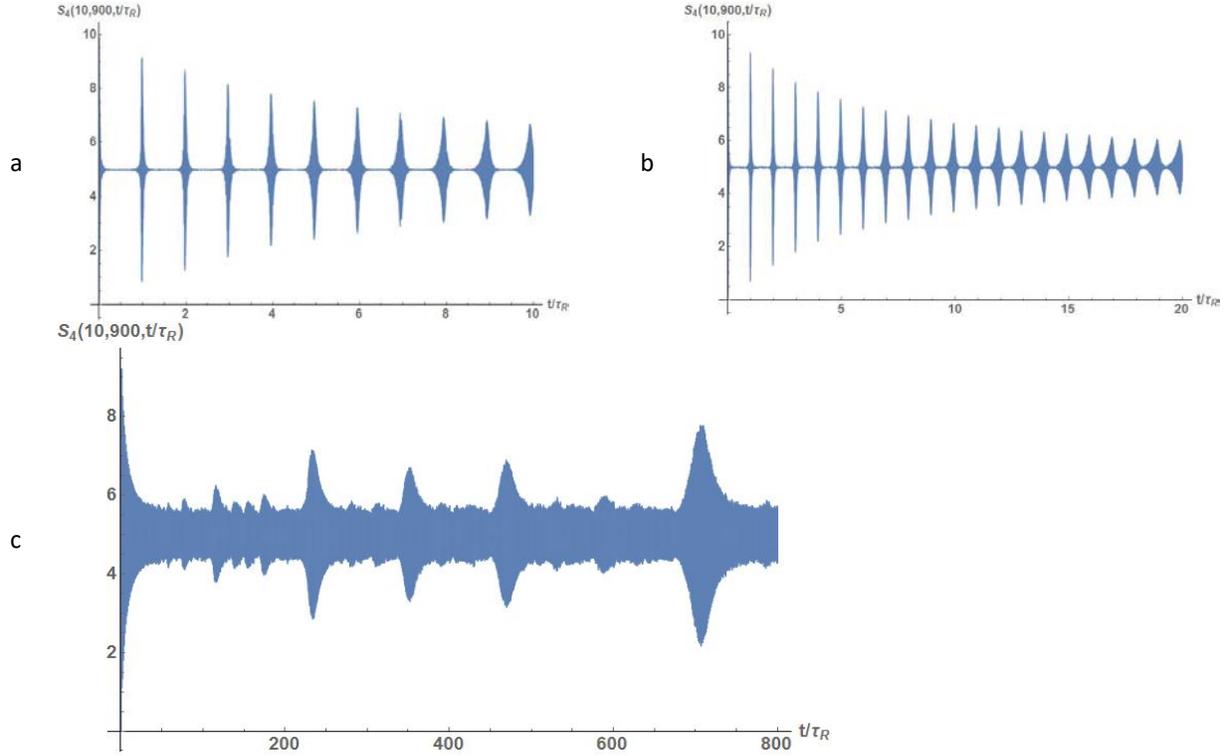

*Fig. 9: Resonant case for the thermal photon distribution with a mean of 10 photons: a) exact case out to a normalized time of 10; b) approximate case out to a normalized time of 20; c) approximate case for normalized time out to 800.*

The behavior is similar to that for the coherent photon distribution; however, the behavior near the second major recovery is not as symmetric nor does the peak amplitude recover to a value of 10.

What has been found is that as normalized time increases, secondary and higher revivals begin to occur and merge into a nearly incoherent display. Of primary interest is that at various times, complete primary revival sets occur with revivals only occurring at integer values of $\tau_R$ such as shown in Figures 6, 7 and 8. Note that this behavior continues for greater times, with the next main peak occurring between 1370 and 1470 $\tau_R$ for N=900, with the peak at 1425 $\tau_R$, which is at twice the value found for the previous set of prime revivals. This second set of prime revivals does not quite reach an amplitude peak of 10. It appears that the location of the prime revival peaks does depend on the number of TLMs. For instance, for 169 TLMs it occurs at 133 $\tau_R$, for 900 TLMs it occurs at 712 $\tau_R$, while for 1600 TLMs the peak occurs at approximately 1269 $\tau_R$ which is the correct ratio of the number of TLMs. Neither the locations of the prime revivals nor the number of revivals seem to be dependent on the mean number of photons.

## 5 A measure of entanglement

Entanglement of TLMs with the quantized field have been discussed by numerous authors in terms of entropy (12, 13, 29, 32-35). In general the authors relate entanglement to the Von Neumann entropy defined as

$$S(\rho) = Tr\rho Ln\rho, \qquad (10)$$

with $\rho$ being the density matrix for the system. However, for a closed system, this quantity should either be zero or a constant. Instead it has been pointed out [29] that for such a system the quantity of interest is the entropy of the reduced density matrix, which is found by taking the trace of $\rho$ over the field or over the TLMs. It was indicated [12, 13] that the entropy of the reduced densities were equal, namely:

$$S(\rho_{TLM}) = Tr\rho_{TLM} Ln\rho_{TLM} = S(\rho_f) = Tr\rho_f Ln\rho_f. \tag{11}$$

For that reason, we focus on $S(\rho_f)$.

This expression may indeed be too complicated to evaluate. Instead, we determine the Shannon reduced entropy [12, 13] of the field defined as

$$\begin{aligned} S_s(\rho_f) &= Tr\rho_f Ln\rho_f \\ &= \sum_{n=0}^{\infty} \langle n|\rho_f|n\rangle \, Ln\langle n|\rho_f|n\rangle. \end{aligned} \tag{12}$$

This is indeed similar to expressions used in [29 and 32] but was called the Von Neumann Entropy. Use the same techniques used to find the initial expression the field ($\langle E^- E^+(t)\rangle$) to obtain the entropy [Appendix G contains the equations for both stimulated emission and stimulated absorption]. The results for stimulated absorption Eq. (G11) are shown below

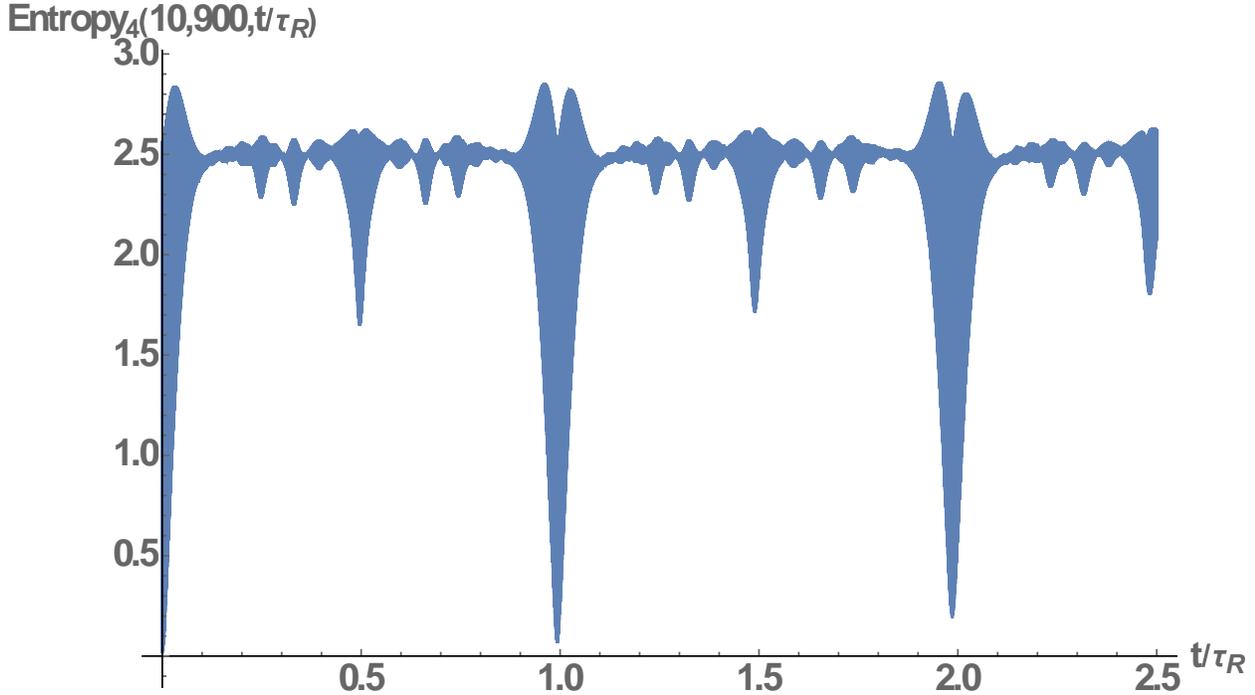

Fig. 10: Shannon entropy for the field for the resonant case with 900 TLMs all originally in the down state and an initial coherent field density matrix with a mean photon number of 10.

The field corresponding to this over the same time range is

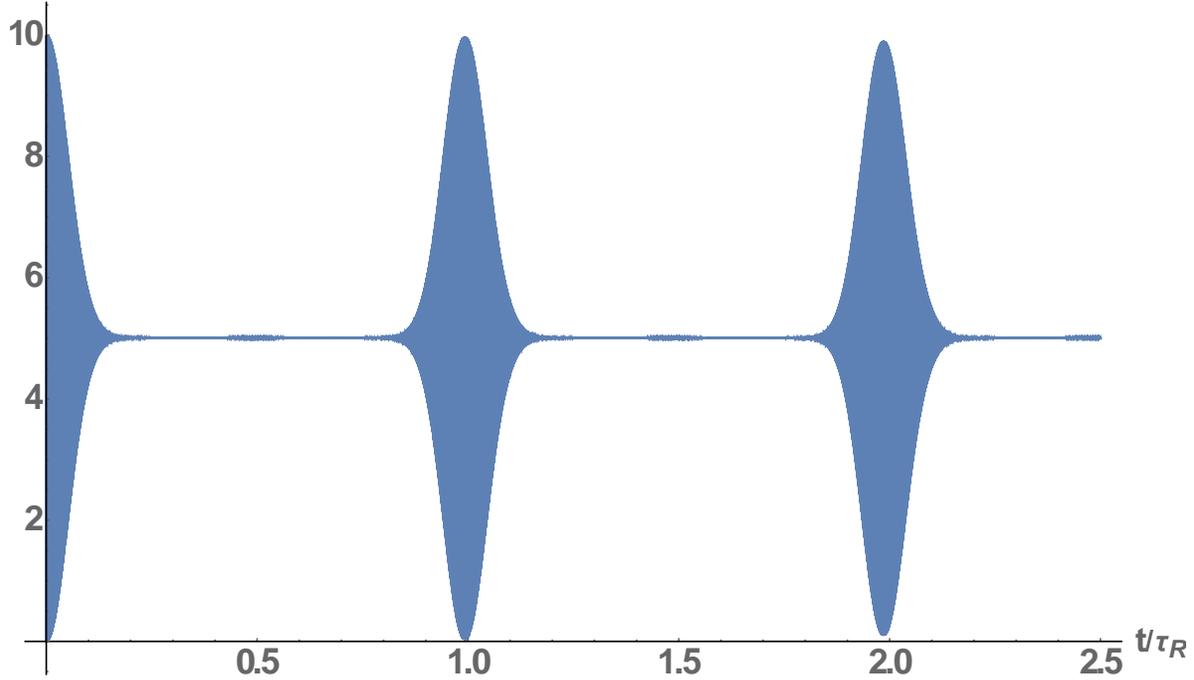

$S_4(10,900,t/\tau_R)$

Figure 11: *Stimulated absorption out to a normalized time of 2 $\tau_R$ for 900 TLMs and a mean photon number of 10*

The calculation time for the field is considerably less that the entropy. Figure 10 indicates that only for integer values of $\tau_R$ does the entropy approach 0, which means near disentanglement for those values. There are other times for which the entropy dips, namely $\tau_R/2$. In addition several small dips are seen between 0 and $\tau_R/2$ and between $\tau_R/2$ and $\tau_R$, which indicates some small amount of disentanglement for those times.

## 6 The Q function

The Q function defined in [36] and extensively used in [37] and [17] as a measure of the time behavior of the coherent state as a function of time. References [17 and 37] address this function for strong fields and small numbers of TLMs. This function can also be addressed for the case here (both stimulated emission and absorption are treated in Appendix G. The results below are for stimulated absorption.), namely:

$$Q(\alpha,t) = \langle \alpha | \rho_f(t) | \alpha \rangle = e^{-|\alpha|^2} \sum_{s=0}^{\infty} \sum_{r=0}^{\infty} \frac{|\alpha|^s |\alpha|^r}{\sqrt{s!\, r!}} e^{-i\varphi(r-s)} < s|\rho_f(t)|r>. \tag{13}$$

The density matrix of the field as a function of time assuming an initial coherent field is given by

$$< s|\rho_f(t)|r> = \sum_{p=0}^{N} \sum_{j=0}^{Min[N,r+p]} \sum_{j'=0}^{Min[N,s+p]} (A^*)^{\frac{N}{2},r-\frac{N}{2}+p,j}_{r+p}$$

$$\times A^{\frac{N}{2},r-\frac{N}{2}+p,j}_{r} e^{-i\lambda_{\frac{N}{2},r-\frac{N}{2}+p,j} t} (A^*)^{\frac{N}{2},s-\frac{N}{2}+p,j'}_{s} A^{\frac{N}{2},s-\frac{N}{2}+p,j}_{s+p} e^{i\lambda_{\frac{N}{2},s-\frac{N}{2}+pj'} t} \langle r+p|\rho_f(0)|s+p \rangle. \tag{14}$$

Again making the assumption that all the TLMs are initially in the down state yields

$$Q(\alpha,t) = e^{-|\alpha|^2-\bar{n}} \sum_{p=0}^{N} \left| \sum_{s=0}^{\infty} \frac{|\alpha|^s \bar{n}^{\frac{s+p}{2}} e^{is\varphi}}{\sqrt{s!(s+p)!}} \sum_{j=0}^{Min[N,s+p]} (A^*)^{\frac{N}{2},s-\frac{N}{2}+p,j'}_{s} A^{\frac{N}{2},s-\frac{N}{2}+p,j'}_{s+p} e^{i\lambda_{\frac{N}{2},s-\frac{N}{2}+p,j'}t} \right|^2. \quad (15)$$

Recall that $\bar{n}$ is the initial mean number of photons. A very good approximation is found to be

$$\widetilde{Q(\alpha,t)} = e^{-|\alpha|^2-\bar{n}} \sum_{p=0}^{N} \left\{ \left| \sum_{s=0}^{\infty} \frac{|\alpha|^s \bar{n}^{\frac{s}{2}} e^{is\varphi}}{s!} \left(\frac{1}{\sqrt{2}}\right)^{s+p} \left[ \sum_{j'=0}^{p} e^{ij'\pi} \sqrt{\frac{(s+p)!}{j'!(s+p-j')!}} \sqrt{\frac{(s+p-j')!p!}{j'!s!}} \frac{F(-s,-j',p-j'+1:-1)}{(p-j')!} \right. \right. \right.$$
$$\times e^{-i4\pi q_{\frac{N}{2},s-\frac{N}{2}+p,j'}\sqrt{N}\frac{t}{\tau_R}}$$
$$+ \sum_{j'=p+1}^{s+p} e^{ij'\pi} \sqrt{\frac{(s+p)!}{j'!(s+p-j')!}} e^{i(j'-p)\pi} \sqrt{\frac{s!j'!}{(s+p-j')!p!}} \frac{F(j'-s-p,-p,j'-p+1:-1)}{(j'-p)!} e^{-i4\pi q_{\frac{N}{2},s-\frac{N}{2}+p,j'}\sqrt{N}\frac{t}{\tau_R}} \right|^2 \right\}. \quad (16)$$

In this equation we have used Eqs (E14-E15) and converted the $\lambda_{\frac{N}{2},s-\frac{N}{2}+p,j'}t$ to $4\pi q_{\frac{N}{2},s-\frac{N}{2}+p,j'}\sqrt{N}t/\tau_R$ where the normalized time is used and resonance was assumed. Examples of the approximate Q function contours are shown in the following table for various values of normalized time. These can be compared to those of ref. [17] for 1, 2 or 3 TLMs. Note that the coordinates are for the real and imaginary parts of $\alpha$.

*Table 6: Approximate Q function as a function of time*

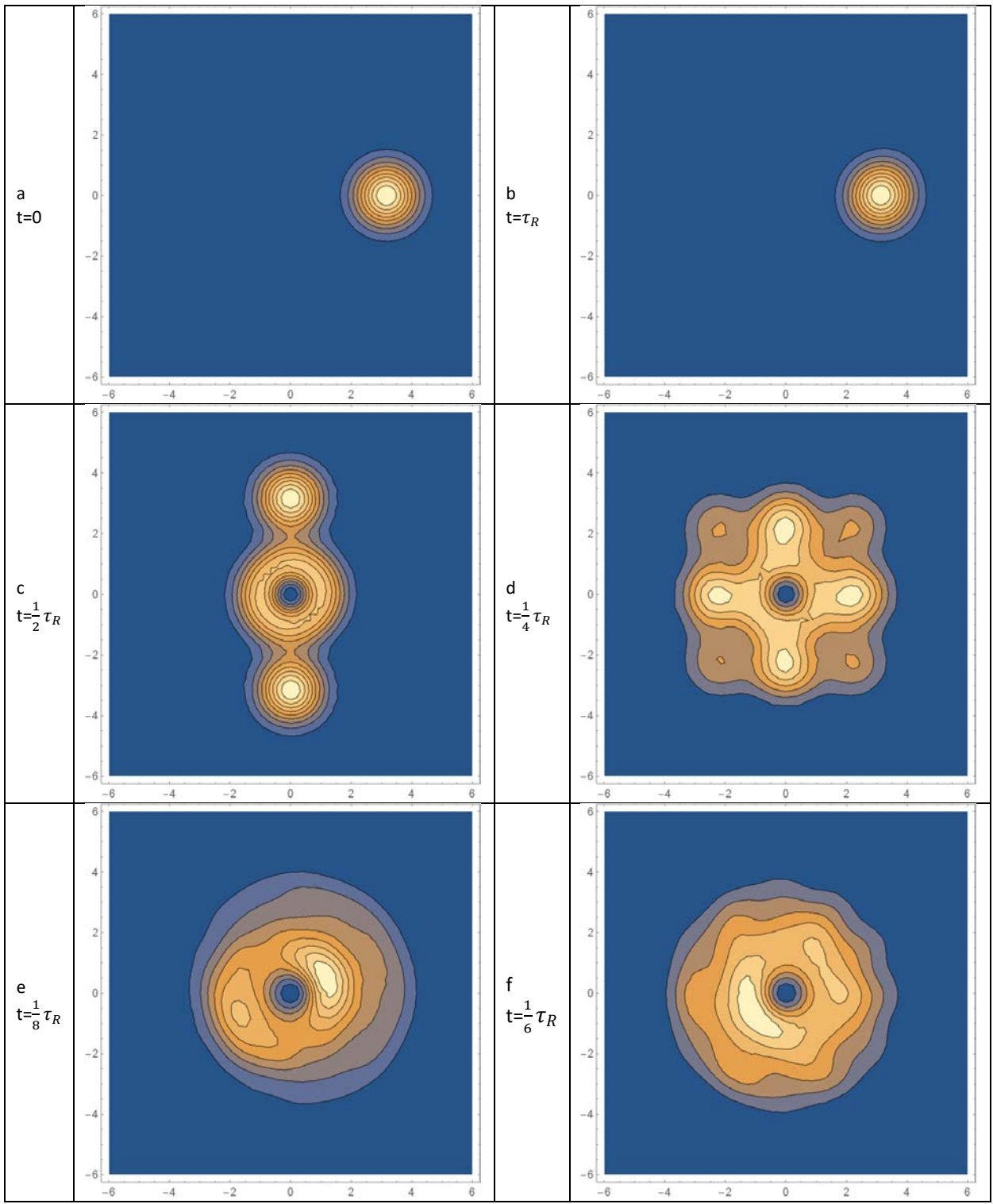

The figures in Table 6 show some interesting behavior. Figure "a" in the table is the coherent distribution at time 0 and indeed shows the peak value for $|\alpha| = \sqrt{\bar{n}}$ as expected. Figure "b" is for $t=\tau_R$ , and the Q function has completely returned to the coherent function, indicating a complete disentanglement of the field and TLMs. This is different than seen previously [17], where complete recovery of the coherent shape was not seen at t=$\tau_R$ or t=$2\tau_R$. The results for the remaining cases in the table are unique. For instance, at $\tau_R/2$ the Q function has split into 3 main lobes, the center lobe centered at $|\alpha|=0$ and having a hollow cone shape. The 2 other lobes are centered at Re $\alpha$=0, Im $\alpha=\pm\sqrt{\bar{n}}$. This is somewhat similar to what was seen in [17] where we did see 2 lobes separate as time increased but for this case we do not necessarily see lobe movement as time increases but we see various configurations based on time. At t=$\tau_R/4$, we see four main lobes and another 4 sub lobes with a center at 0. Each of the 4 main lobes are relatively symmetric with centers at Re $\alpha$=0, Im $\alpha=\pm 2$ or Im $\alpha$=0, Re $\alpha=\pm 2$. The sub lobes are centered at Re $\alpha=\pm 2$, Im. $\alpha=\pm 2$ but have a considerable smaller amplitude. The other 2 examples for t=/8 and t=/6 show the Q function as not well defined and spread out over a rather large area and thus of smaller amplitude. This implies that the Q function has relatively well-defined structure for times associated with integer values of $\pi$ with complete reconstitution only for values which are integer values of the revival time. Note that the calculations used to generate the figures in Table 4 were only carried out to a TLM value of 30. It is believed that this is of sufficient accuracy. To check this assumption the calculation was carried out to N=60 for t=/4 and t=/6 with no significant differences.

**7 Summary**

In this paper we have provided the time dependent part of the exact (Eq. (5)) and approximate equations (Eqs. (6) and (6a)) for the time development for field intensity in terms of the eigenvalues and eigenvalues of the Tavis-Cumming model (Hamiltonian). We find that for large numbers of TLMs initially in the down state (big negative spin) all at equivalent mode positions within a closed lossless cavity interacting with an initial photon distribution with a SMPN, persistent photon echoes are seen with revival times proportional to the square root of the number of TLMs. This new characteristic time is not the same as for the case of small numbers of TLMs interacting with a field with a LMPN. It is also found that the initial revivals and collapse behavior decays at times proportional to the square root of the number of TLMs times the revival time by which time secondary and tertiary revivals begin to grow. In addition we have found an equivalent Rabi time inversely proportional to the square root on the TLM number and a collapse time equal to $\tau_R/4\pi$. At much larger times proportional to the number of TLMs times the revival time and at multiples thereof, complete revival sets of the primary revivals are seen. These new revival sets also slowly decrease in amplitude for the second and higher multiple times.

Most of the numerical examples were for the coherent photon distribution. However, within this paper we also considered an initial thermal distribution which displays the same persistent echo behavior including the longer time behavior seen for the coherent case. The complete revival at long times is not as complete as for the coherent case. We have also found elsewhere, that the revival time is independent of the initial photon distribution and proportional to the square root of the number of TLMs except in the case where the photon distribution does not have a valid value for each photon number.

To examine the question of entanglement and disentanglement we have examined the entropy of the density matrix of the field. We have not found disentanglement of the TLMs and field except at multiples of the revival time $\tau_R$ where the entropy drops to 0. At intermediate times there is some minor disentanglement but it is not as significant as for low numbers of TLMs and high mean photon number, especially for the cases where the TLM state is started in the so-called attractor state.

Finally we have examined the Q function for the case of the initial photon distribution being the coherent state. This function specifies the probability distribution of the coherent state as a function of time. We find that this function returns to the initial photon distribution at the revival times. Interesting cases exist when the time is equal to $\tau_R/2$ or $\tau_R/4$ or when the time is at integer multiples of $\pi$. At other times this function appears ill defined but is probably due to the interaction of many mini states of the coherent state, one for each TLM each moving at different velocities around the complex $\alpha$ plain. For a much smaller number of TLMs this behavior was distinctly observed by other authors [17, 37].

It is intriguing to see such long-term persistence of revivals and then the long-term re-emergence of revival sets. It may possibly be argued that the results are not physical and are due strictly to a model which does not include dissipation or a driving function. However, the same could have been said about the behavior seen in [7,8] until the experimental results in [11] was presented. It should be noted, that the absolute values of the characteristic times discussed in this paper depend on the value of the coupling constant γ. In refs [1, 3 and 7] this coupling constant was taken as 5 cps which leads to an absolute revival time of 23 seconds for an ammonia beam maser. The other characteristic times are equally large and appear unphysical when compared to the Q of a cavity. On the other hand, Gea-Banacloche, [15] has postulated a coupling constant of 276 kHz which leads to a revival time of .001 sec for 900 TLMs. This is much more physically plausible except that the model was for a micromaser and a small cavity which leads to the question of accommodation for a large number of TLMs. This leads to a dilemma on which coupling constant to choose (or to explain the dilemma) or the need to carefully define an experiment appropriate to large numbers of TLMs and few photons and an associated coupling constant. Perhaps this work will lead to new physical insights in systems such as quantum computing with alternate structures than have been considered previously.

**Appendices: Mathematical Background for Photon Echoes**

The set of appendices below provide a nearly complete set of derivations need for the study of a system of many (few) two-level molecules (TLMs) all interacting with a photon field in a closed cavity. We begin with the exact solution to the TCM Hamiltonian with definitions of the various constants and the expression of the eigenvectors in terms of the TLM and photon states. Then the general solution of the time development of $E^-E^+$ (photon echoes) in terms of the eigenvectors of the TCM Hamiltonian is developed. From the general expression specific equations for stimulated emission with the TLMs all initially in the up state and for stimulated absorption with all the TLMs initially in the down state are found. Note that there all solutions for the case of half the TLMs in the up state and half in the down state and also for a few TLMs in the up state while the rest are in the down state. Note also that the derivations are based on the viewpoint of the photon field rather that the viewpoint of the TLMs (which has been considered much more often).

Approximate solutions for stimulated emission and absorption are then developed in terms of the approximate eigenvalues of the TCM Hamiltonian. Entropy and the Q function of the photon density matrix are also considered.

### APPENDIX A: THE SOLUTION TO THE TCM HAMILTONIAN

The Hamiltonian which describes the interaction between N identical TLMs at equivalent mode positions interacting with a quantized radiation field in a closed lossless cavity using the rotating wave approximation [3] is given by

$$H = (a^\dagger a + R_3) + a^\dagger a \frac{\omega - \Omega}{\Omega} - \kappa a R_+ \quad \text{(A1)}$$
$$- \kappa^* a^\dagger R_-$$

$$\kappa = \frac{\gamma}{\Omega}. \quad \text{(A2)}$$

The energy separation of the TLM is $\Omega$, the photon energy is $\omega$ and $\gamma$ is the complex coupling between the photons and the TLMs. To simplify the equation, Plank's constant divided by $2\pi$ ($\hbar$) has been set equal to unity. Note that $\Omega$ is not equal to $\omega$ indicating that this is the Hamiltonian for non-resonance. For the resonant case, the two are set equal.

States of the non-interacting system are defined such that

$$H_o |n> |r,m> = \left(m + n + n\frac{\omega - \Omega}{\Omega}\right) |n> |r,m> \tag{A3}$$

$$R_3 |r,m> = \sum_{j=1}^{N} R_{j3} |r,m> = m |r,m> \tag{A4}$$

$$R_\pm |r,m> = \sum_{j=1}^{N} R_{j\pm} |r,m> = e^{\pm i\varphi_2}(r(r+1) - m(m\pm 1))^{\frac{1}{2}} |r,m> \tag{A5}$$

and

$$a|n> = e^{i\varphi_3} \sqrt{n} |n-1> \tag{A6}$$

The states $|r,m>$ are formally identical to states of total angular momentum and total z-component of momentum for a system of spins. These $|r,m>$ states are formed in the same way, from single TLM states $|1/2, \pm 1/2>$, as the total spin states are formed from the individual spinor states. The "cooperation number," r, analogous to the total angular momentum of a spin system, satisfies

$$R^2 |r,m> = r(r+1) |r,m>, \tag{A7}$$

with m$\leq r \leq \frac{N}{2}$ where r and m are either integer of half-integer. The term "cooperation number" has significance in the fact that the projection of the vector operator on the "1-2" plane gives essentially the total dipole moment of the TLM system while the third component, $R_3$, gives the energy. For larger "cooperation number," the possible interaction with the electromagnetic field also becomes larger. On the other hand, a value of r = 0 implies no interaction with the electromagnetic field at all. The operators in Eq. (A1) satisfy the commutation relations

$$[R_3, R_\pm] = \pm R_\pm \tag{A8}$$

$$[R_+, R_-] = 2R_3 \tag{A9}$$

$$[a, a^\dagger] = 1. \tag{A10}$$

Since both $R^2$ and c = $a^\dagger a + R_3$ commute with H, the eigenstates may be chosen to be eigenstates of these two operators as well, and the eigenstates may be labeled by the eigenvalues r and c. The symbol c represents the conservation of the number of the photons plus the number of TLMs in the excited state. If $\omega=\Omega$, c is the eigenvalue of $H_o$ and is given by Eq. (A3). For a given r and c, there will be (in general) 2r+1 energy eigenvalues and for ω=Ω these eigenvalues will be symmetrically displaced about the constant c, where -r < c < ∞ .

Denote the eigenstates of H as |r, c, j > .

$$H|r,c,j> = \lambda_{r,c,j}|r,c,j> \tag{A11}$$

where j takes on the 2r + 1 values 0,1,2, . . . , 2r, if c ≥ r, or the c + r + 1 values 0,1,⋯,c+r if c < r.

Recalling that the states |r, c, j > are eigenstates of H, c, and R², and that m = c-n varies between r and -r,

$$|r,c,j> = \sum_{n=max(0,c-r)}^{c+r} A_n^{(r,c,j)}|n>|r,c-n> \tag{A12}$$

The $A_n^{(r,c,j)}$ satisfy the difference equation

$$-|\kappa|e^{-i\varphi}\sqrt{n}C_{r,c-n}A_{n-1}^{(r,c,j)} + \left(c + \frac{\omega-\Omega}{\Omega}n - \lambda_{r,c,j}\right)A_n^{(r,c,j)} \\ - |\kappa|e^{i\varphi}\sqrt{n+1}C_{r,c-n-1}A_{n+1}^{(r,c,j)} = 0, \tag{A13}$$

where

$$\varphi = \varphi_1 + \varphi_2 + \varphi_3, \tag{A14}$$

and

$$C_{r,c-n} = [r(r+1) - (c-n)(c-n+1)]^{1/2}. \tag{A15}$$

The A⁽ʳ ᶜ ʲ⁾ satisfy the boundary conditions

$$A_{r+c+1}^{(r,c,j)} = A_{max(-1,c-r-1)}^{(r,c,j)} = 0. \tag{A16}$$

It is convenient to define $B_n$'s so that

$$A_n = \frac{(e^{-i\varphi})^n B_n}{\sqrt{n!}\, C_{min(c-1,r-1)}C_{min(c-2,r-2)} \cdots C_{c-n}} \tag{A17}$$

The superscripts (subscripts) (r, c, j) have been dropped for simplicity whenever this does not cause confusion. Only the r subscript will be suppressed in the $C_{r,c-n}$. If an effective eigenvalue

$$q = \frac{c - \lambda}{|\kappa|} \tag{A18}$$

and relative tuning parameter

$$\beta = \frac{\omega - \Omega}{|\kappa|\Omega} \tag{A19}$$

are defined, $B_n$ will satisfy the difference equation

$$B_{n+1} - (q + \beta n)B_n + nC_{c-n}^2 B_{n-1} = 0. \tag{A20}$$

The largest value of q for a given value of r and c corresponds to the ground state of the system.

The exact solution of Eq. (A20) (non-normalized) can be obtained by "unraveling" from one end. That is, by starting from

$$B_{max(-1,c-r-1)} = B_{c+r+1} = 0, \tag{A21}$$

one may obtain a solution for $B_n$ in the form

$$B_n = \sum_{\ell=0}^{[t/2]} (-1)^\ell S_\ell^{(t-1)}, \tag{A22}$$

where

$$n = t + \alpha \tag{A23}$$

$$\alpha = max(0, c - r), \tag{A24}$$

and $[t/2]$ is the first integer equal to or less than t/2.

The $S_\ell^t$ are given by

$$S_\ell^t = \sum_{m_1=1}^{t} \sum_{m_2=m_1+2}^{t} \sum_{m_3=m_2+2}^{t} \cdots \sum_{m_\ell=m_{\ell-1}+2}^{t} \left\{ \prod_{\substack{y=0 \\ y \neq [m_i, m_i-1]}}^{t} (q + (y + \alpha)\beta) \right\} \prod_{j=\{m_i\}} C_j, \tag{A25}$$

where $C_{m_i} = (m_i + \alpha)C^2_{c-(m_i+\alpha)}$. In Eq. (A25) the first product cannot contain terms with y equal to any of the $m_i$ or $m_i$-1 and the second product contains only terms with j equal to one of the $m_i$.[2]

The above equation (A25) is much too difficult to use in a practical way; however, a recursion relation for the "$S_\ell^t$" exists which makes the use of Eq. (A22) practical. This recursion relation

$$S_\ell^t = (q + (\alpha + t)\beta)S_\ell^{t-1} + C_t S_{\ell-1}^{t-2} \tag{A26}$$

is found by induction or inspection.

The exact eigenvalues, or equivalently the q's, are determined from Eq. (22) for $B_{r+c+1}$ namely

$$B_{r+c+1} = \sum_{\ell=0}^{[t'/2]} (-1)^\ell S_\ell^{(t'-1)} = 0, \tag{A27}$$

where t' = r+c+1-$\alpha$. These are polynomials in q of degree 2r+1 if c≥ r, and of degree r+c+l if c<r. If 2r+1 or r+c+1 is even and $\beta$=0, one of the roots of Eq. (A27) is q=0. For this case, the eigenvector can be found directly from the equation for the $A_n$ and is given by

$$A_n = (-1)^{t/2} e^{-it\varphi} \sqrt{\frac{C_{t-1}C_{t-3}\cdots C_1}{C_t C_{t-2}\cdots C_2}}, \; t \text{ even}$$
$$= 0 \qquad\qquad\qquad\qquad , t \text{ odd} \tag{A28}$$

Also in the special case ($\beta = 0$) the q's are such that $q_{2r}$=-$q_0$, $q_{2r-1}$=-$q_1$, and so on symmetrically placing the q-values about zero, and the states $A_n^{(j)}$, j =2r, 2r-1,$\cdots$, r-1, if r is an integer (or r+1/2 if r is a half-integer) are found from the states $A_n$, j = 0, 1,$\cdots$, r-1 (or r-1/2 if r is a half-integer) by replacing $\varphi$ by $\varphi + \pi$ in Eq. (A17). Care must be taken with these latter statements since $0 \leq j \leq r + c$ when c < r.

This completes the discussion of the determination of the exact eigenvectors and eigenvalues of the Hamiltonian. Evaluation of the eigenvalues and eigenvectors is carried out using Mathematica rather than the techniques in [3]. The Mathematic code used for evaluation is provided below.

ab[nM_,n_,β_]:=Block[{α,r,c,n1,max,nmax,m,h,g,nn,b,sols,vals,bn,nrm1,ba,an}
Comment: The variables in the {} brackets are defined as local variables.
,r=nM/2;c=n±r;α=Max[0,c-r];n1=c+r;max=n1-α+1;nmax=Floor[(max+1)/2]
Comment: We are defining the value of variables above.
;h[k_]:=q+(k+α)β;g[k_]:=(r (r+1)-(c-k-α)(c-k-α+1)) (k+α);nn=Table[(-1)^(j-1),{i,1,max+1},{j,1,nmax+1}];m=Table[0,{i,0,nmax},{j,0,max}];m[[1,1]]=1;Do[m[[1,j]]=m[[1,j-1]] h[j-2],{j,2,max+1}];Do[m[[i,j]]=h[j-2] m[[i,j-1]]+g[j-2] m[[i-1,j-2]],{i,2,nmax+1},{j,2 i-1,max+1}];m=Transpose[m];m=m nn;b=Plus@@@m;sols=NSolve[b[[max+1]]==0,q,WorkingPrecision->60];b=Drop[b,-1];vals=Sort[Re[Table[q/.sols[[j]],{j,1,max}]],#1>#2 &];bn=Table[b/.q->vals[[j]],{j,1,max}];nrm1=Range[max];nrm1[[1]]=Sqrt[(n1-α)!/(α! (r-c+α)!)];Do[nrm1[[j]]=nrm1[[j-1]]/Sqrt[(α+j-1)(r-c+α+j-1)(r+c-α-j+2)],{j,2,max}];bn=Transpose[Transpose[bn] nrm1];ba=Table[Norm[bn[[j]]],{j,1,max}];an=bn/ba;

---
[2] Note that Eq. (A25) consists of the sum of products of $C_j$ taken $\ell$ at a time with no nearest neighbors.

{vals, an, max-1}];

This is a functional definition with the 3 input values for the number of TLMs, the number of photons and β as defined in Eq. (A19). A block command is used to perform all the calculations using local variables. Note that c=n±r depending if all the TLM are initially all in the up state or down state. The g function is the evaluation of $C_{m_i}$. The h function is one of the factors in Eq. (A26). The table m is the matrix of the values of $S_\ell^t$ using the recursion relation (A26). We have added the sign function for the calculation of (A22). Eq. (A27) is used to form the equation necessary to find the eigenvalues (vals) in the above expression. . The Bs are then found and normalized. Finally the As (A17) are defined.

We next discuss the time development of $E^-E^+$ using the eigenvectors and eigenvalues found above.

## APPENDIX B: ENSEMBLE AVERAGES FOR $E^-E^+$

Unlike others exploring this field, we focus on the time development of the photon distribution rather the density matrix for the TLM state. Then the ensemble average of the field operator E⁻E⁺(t) may be found in the usual way [27]

$$\langle E^-E^+(t)\rangle = \left|\frac{\gamma}{\mu}\right|^2 \sum_{n=0}^{\infty} n \langle n|\rho_f(t)|n\rangle, \tag{B1}$$

where only one mode of the field is excited, $\gamma$ is the complex coupling constant, and $\mu$ the dipole moment of the TLMs with which the field is interacting. The element of the field density matrix is given by the trace over the TLM states

$$\langle n|\rho_f(t)|n'\rangle = \sum_{r,m} P(r) \langle n|\langle r,m|\rho(t)|r,m\rangle|n'\rangle, \tag{B2}$$

where

$$P(r) = \frac{N!\,(2r+1)}{\left(\frac{N}{2}+r+1\right)!\left(\frac{N}{2}-r\right)!} \tag{B3}$$

and $\rho(t)$ is given by a unitary transformation of the density operator at time $t_o=0$ where it is assumed that the N-TLMs and radiation field are not interacting! Therefore

$$\rho(t) = U(t)\rho(0)U^{-1}(t), \tag{B4}$$

$$U(t) = e^{iHt}, \tag{B5}$$

and where H is given above
Since the system is non-interacting at time zero, the density operator is a direct product of the field part and N-TLM part of the system.

$$\rho(0) = \rho_{TLM} \otimes \rho_f. \tag{B6}$$

The expression above for E⁻E⁺(t) may be expanded in terms of the eigenvectors of the Hamiltonian or in terms of the |r, m>|n> states due to orthogonality and completeness. Further note that

$$|n\rangle |r,c-n\rangle = \sum_{j=0}^{Min[2r,c+r]} (A^*)_n^{r,c,j} |r,c,j\rangle. \tag{B7}$$

In addition

$$U(t,0)|n\rangle |r,c-n\rangle = \sum_{j=0}^{Min[2r,c+r]} (A^*)_n^{r,c,j} e^{-i\lambda_{r,c,j}t} |r,c,j\rangle, \tag{B8}$$

With this, equation (B1) can be written as[3]

$$\langle E^-E^+(t)\rangle = \left|\frac{\gamma}{\mu}\right|^2 \sum_{n=0}^{\infty} n\langle n|\rho_f(t)|n\rangle$$

$$= \left|\frac{\gamma}{\mu}\right|^2 \sum_{r=0,\frac{1}{2}}^{\frac{N}{2}} P(r) \sum_{m=-r}^{r} \sum_{c=m}^{\infty} (c-m) \sum_{p=-r}^{r} \sum_{j=0}^{Min[2r,c+r]} \sum_{p'=-r}^{r} \sum_{j'=0}^{Min[2r,c+r]} (A^*)_{c-p}^{r,c,j} \tag{B9}$$

$$\times A_{c-m}^{r,c,j}(A^*)_{c-m}^{r,c,j'} A_{c-p'}^{r,c,j'} e^{-i\lambda_{r,c,j}t} e^{i\lambda_{r,c,j'}t} \langle c-p|\rho_f(0)|c-p'\rangle \langle r,p|\rho_{TLM}(0)|r,p'\rangle,$$

As is seen, this general equation is quite complicated and the use of the general expression would is prohibitive except for the simplest of cases. This completes the formulation and review of the general expression. Two cases (there are others) which simplify the equation are all TLMs initially in the up state and all TLMs initially in the down state. For stimulated emission and all TLMs in the up state $\langle E^-E^+(t)\rangle$ is given by

$$\langle E^-E^+(t)\rangle = \left|\frac{\gamma}{\mu}\right|^2 [\bar{n} + S_1(\bar{n},N,\gamma t)], \tag{B10}$$

where

$$S_1(\bar{n},N,\gamma t) = -4 \sum_{n=0}^{\infty} \langle n|\rho_f(0)|n\rangle \sum_{j=0}^{N-1} \sum_{j'=j+1}^{N} Sin^2\left\{\frac{\left[q_{\frac{N}{2},(n+\frac{N}{2}),j} - q_{\frac{N}{2},(n+\frac{N}{2}),j'}\right]}{2}\Omega|\kappa|t\right\}$$

$$\times (A^*)_n^{\frac{N}{2},(n+\frac{N}{2}),j} A_n^{\frac{N}{2},(n+\frac{N}{2}),j'} \sum_{p=0}^{N} p A_{n+p}^{\frac{N}{2},(n+\frac{N}{2}),j} (A^*)_{n+p}^{\frac{N}{2},(n+\frac{N}{2}),j'}. \tag{B11}$$

Note that we have used Eq. (A18) to relate $\lambda$ to q and multiplied $\Omega$ to obtain the correct units. For stimulated absorption and all TLMs in the down state

$$\langle E^-E^+(t)\rangle = \left|\frac{\gamma}{\mu}\right|^2 [\bar{n} - S_4(\bar{n},N,\gamma t)], \tag{B12}$$

where

---

[3] The factors $\left(\frac{\gamma}{\mu}\right)$ and $\left|\frac{\gamma}{\mu}\right|^2$ seen in eq. (37) differ from those same constants in Reference 1 by a factor of 2, i.e. $\gamma$ is $2\gamma$ in that reference. We ignore the difference here.

$$S_4(\bar{n}, N, \gamma t) = -4 \sum_{n=0}^{\infty} \langle n|\rho_f(0)|n\rangle \sum_{j=0}^{Min[N,n]-1} \sum_{j'=j+1}^{Min[N,n]} Sin^2 \left\{ \frac{\left[q_{\frac{N}{2},\left(n-\frac{N}{2}\right),j} - q_{\frac{N}{2},\left(n-\frac{N}{2}\right),j'}\right]}{2} \Omega |\kappa| t \right\}$$
$$\times (A^*)_n^{\frac{N}{2},n-\frac{N}{2},j} A_n^{\frac{N}{2},n-\frac{N}{2},j'} \sum_{p=0}^{min[N,n]} pA_{n-p}^{\frac{N}{2},n-\frac{N}{2},j} (A^*)_{n-p}^{\frac{N}{2},n-\frac{N}{2},j'}$$
(B13)

Note that $\bar{n}$ is the mean number of photons and that the notation of 1 or 4 on the subscripts of S correspond to that used by Cummings [7]. Slightly different forms for $S_1$ and $S_4$ which may make the calculation a little easier can be found. For instance an alternate form for $S_4$ is given by

$$S_4(\bar{n}, N, \gamma t) = \sum_{n=0}^{\infty} \langle n|\rho_f(0)|n\rangle \sum_{p=0}^{Min[N,n]} p \left\{ \left| \sum_{j=0}^{Min[N,n]} (A^*)_n^{\frac{N}{2},n-\frac{N}{2},j} A_{n-p}^{\frac{N}{2},n-\frac{N}{2},j} Cos\left(q_{\frac{N}{2},n-\frac{N}{2},j}\Omega|\kappa|t\right) \right|^2 \right.$$
$$\left. + \left| \sum_{j=0}^{Min[N,n]} (A^*)_n^{\frac{N}{2},n-\frac{N}{2},j} A_{n-p}^{\frac{N}{2},n-\frac{N}{2},j} Sin\left(q_{\frac{N}{2},n-\frac{N}{2},j}\Omega|\kappa|t\right) \right|^2 \right\}$$
(B14)

Equations (B11) as well as (B13-B14) can be modified slightly by using normalized values of time. This normalization will make the equations more complicated but provide graphical representations of those equation display photon revivals occurring at integer values of the normalized time. For example, Eberly [4] defined a normalized time for the TLMs in the up state as

$$\tau = \frac{t}{\tau_R} \text{ where } 2\pi \frac{\sqrt{\bar{n}+1+\Delta}}{\gamma},$$
(B15)

where $\Delta = \beta^2/4$. Similarly, for the TLMs initially in the down state using the same notation

$$\tau = \frac{t}{\tau_R} \text{ where } 4\pi \frac{\sqrt{N+\Delta}}{\gamma},$$
(B16)

where N is the number of TLMs. Using these normalizations $S_1$ and $S_4$ can be rewritten as

$$S_1(\bar{n}, N, \gamma t) = -4 \sum_{n=0}^{\infty} \langle n|\rho_f(0)|n\rangle \sum_{j=0}^{Min[N,n]-1} \sum_{j'=j+1}^{Min[N,n]} G(\tau)(A^*)_n^{\frac{N}{2},\left(n+\frac{N}{2}\right),j} A_n^{\frac{N}{2},\left(n+\frac{N}{2}\right),j'} \times \sum_{p=0}^{N} p A_{n+p}^{\frac{N}{2},\left(n+\frac{N}{2}\right),j} (A^*)_{n+p}^{\frac{N}{2},\left(n+\frac{N}{2}\right),j'}$$
(B17)

$$S_4(\bar{n}, N, \gamma t) = -4 \sum_{n=0}^{\infty} \langle n|\rho_f(0)|n\rangle \sum_{j=0}^{Min[N,n]-1} \sum_{j'=j+1}^{Min[N,n]} H(\tau)(A^*)_n^{\frac{N}{2},n-\frac{N}{2},j} A_n^{\frac{N}{2},n-\frac{N}{2},j'} \times \sum_{p=0}^{min[N,n]} pA_{n-p}^{\frac{N}{2},n-\frac{N}{2},j} (A^*)_{n-p}^{\frac{N}{2},n-\frac{N}{2},j'}$$
(B18)

$$G(\tau) = Sin^2\left\{\pi\sqrt{\bar{n}+1+\Delta}\left[q_{\frac{N}{2},\left(n+\frac{N}{2}\right),j} - q_{\frac{N}{2},\left(n+\frac{N}{2}\right),j'}\right]\tau\right\}$$
(B19)

$$H(\tau) = Sin^2\left\{2\pi\sqrt{N+\Delta}\left[q_{\frac{N}{2},\left(n-\frac{N}{2}\right),j} - q_{\frac{N}{2},\left(n-\frac{N}{2}\right),j'}\right]\tau\right\}$$
(B20)

### **APPENDIX C: SOME SIMPLE EXAMPLES**
In this appendix the exact solutions for small numbers of TLMs are presented for reference

## 1 TLM
The expression for the density matrix for the field for 1 TLM in the up state and non-resonance is given by

$$S_1\left(\bar{n}, 1, \frac{t}{\tau_R}, \Delta\right)_{NR} = \sum_{n=0}^{\infty} \langle n|\rho_f(0)|n\rangle \left(\frac{n+1}{n+1+\Delta}\right) Sin^2[2\pi\sqrt{\bar{n}+1+\Delta}\sqrt{n+1+\Delta}\tau]. \qquad (C1)$$

Note that the subscript NR means that the results are for non-resonance. Other simple exact results for 2, 3 and 4 TLMS initially in the up state for the resonant case are provided below.

## 2 TLMs
For N=2 and $\beta = 0$, $S_1(\bar{n}, 2, \gamma t)$ is given by;

$$\begin{aligned}S_1(\bar{n}, 2, \gamma t) = 8 \sum_{n=0}^{\infty} \langle n|\rho_f(0)|n\rangle &\left\{\frac{(n+1)(n+2)}{(2n+3)^2} Sin^2\left[2\pi\sqrt{\bar{n}+2}\sqrt{n+\frac{3}{2}}\tau\right]\right. \\ &\left. - \frac{1}{8}\frac{(n+1)}{(2n+3)^2} Sin^2\left[4\pi\sqrt{\bar{n}+2}\sqrt{n+\frac{3}{2}}\tau\right]\right\}\end{aligned} \qquad (C2)$$

## 3 TLMs
For N=3 and $\beta = 0$, $S_1(\bar{n}, 3, \gamma t)$ is given by;

$$\begin{aligned}S_1(\bar{n}, 3, \gamma t) = 4 \sum_{n=0}^{\infty} \langle n|\rho_f(0)|n\rangle &\left\{\frac{3(2+n)(1+n+\sqrt{(1+n)(3+n)})}{73+16n(4+n)}\right. \\ &\times Sin\left[2\pi\sqrt{\bar{n}+3}\left(\sqrt{10+5n-\sqrt{73+16n(4+n)}} - \sqrt{10+5n+\sqrt{73+16n(4+n)}}\right)\tau\right]^2 \\ &+ \frac{3(1+n)(2+n)(8+4n+\sqrt{73+16n(4+n)})}{2(73+16n(4+n))(-7-2n+\sqrt{73+16n(4+n)})} Sin\left[4\pi\sqrt{\bar{n}+3}\sqrt{10+5n-\sqrt{73+16n(4+n)}}\tau\right]^2 \\ &- \frac{3(2+n)(-1-n+\sqrt{(1+n)(3+n)})}{73+16n(4+n)} \\ &\times Sin\left[2\pi\sqrt{\bar{n}+3}\left(\sqrt{10+5n-\sqrt{73+16n(4+n)}} + \sqrt{10+5n+\sqrt{73+16n(4+n)}}\right)\tau\right]^2 \\ &\left. + \frac{3(1+n)(2+n)(-8-4n+\sqrt{73+16n(4+n)})}{2(73+16n(4+n))(7+2n+\sqrt{73+16n(4+n)})} Sin\left[4\pi\sqrt{\bar{n}+3}\sqrt{10+5n+\sqrt{73+16n(4+n)}}\tau\right]^2\right\}\end{aligned} \qquad (C3)$$

## 4 TLMs
For N=4 and $\beta = 0$, $S_1(\bar{n}, 4, \gamma t)$ is given by;

$$S_1(\bar{n}, 4, \gamma t) = 4 \sum_{n=0}^{\infty} \langle n|\rho_f(0)|n\rangle$$

$$\times \left\{ Z1(n) \mathrm{Sin}\left[2\pi\sqrt{n+4}\left(\sqrt{25+10n-3\sqrt{33+4n(5+n)}} - \sqrt{25+10n+3\sqrt{33+4n(5+n)}}\right)\tau\right]^2 \right.$$

$$+ Z2(n) \mathrm{Sin}\left[2\pi\sqrt{n+4}\sqrt{25+10n-3\sqrt{33+4n(5+n)}}\,\tau\right]^2 - Z3(n) \mathrm{Sin}\left[2\pi\sqrt{n+4}\sqrt{25+10n+3\sqrt{33+4n(5+n)}}\,\tau\right]^2$$

$$- Z4(4) \mathrm{Sin}\left[4\pi\sqrt{n+4}\sqrt{25+10n-3\sqrt{33+4n(5+n)}}\,\tau\right]^2$$

$$- Z5(5) \mathrm{Sin}\left[2\pi\sqrt{n+4}\left(\sqrt{25+10n-3\sqrt{33+4n(5+n)}} + \sqrt{25+10n+3\sqrt{33+4n(5+n)}}\right)\tau\right]^2$$

$$\left. + Z6(n) \mathrm{Sin}[4\pi\sqrt{n+4}\sqrt{25+10n+3\sqrt{33+4n(5+n)}}\,\tau]^2 \right\} \quad (C4)$$

In Eq. {C4)

$$Z1(n) = \frac{(1+n)(2+n)(3+n)(10+4n+\sqrt{82+16n(5+n)})}{(33+4n(5+n))(41+8n(5+n))}$$

$$Z2(n) = \frac{12(1+n)(2+n)(3+n)(4+n)(1+2n+\sqrt{33+4n(5+n)})}{(41+8n(5+n))(561-87\sqrt{33+4n(5+n)}+n(505-45\sqrt{33+4n(5+n)}+2n(84+10n-3\sqrt{33+4n(5+n)})))}$$

$$Z3(n) = \frac{12(-1-2n+\sqrt{33+4n(5+n)})\mathrm{Gamma}[5+n]}{(41+8n(5+n))(561+87\sqrt{33+4n(5+n)}+n(505+20n^2+45\sqrt{33+4n(5+n)}+6n(28+\sqrt{33+4n(5+n)})))n!}$$

$$Z4(n) = \frac{3(1+n)(2+n)(3+n)(9+\sqrt{33+4n(5+n)}+n(7+2n+\sqrt{33+4n(5+n)}))}{(561-87\sqrt{33+4n(5+n)}+n(505-45\sqrt{33+4n(5+n)}+2n(84+10n-3\sqrt{33+4n(5+n)})))^2}$$

$$Z5(n) = \frac{(1+n)(2+n)(3+n)(-10-4n+\sqrt{82+16n(5+n)})}{(33+4n(5+n))(41+8n(5+n))}$$

$$Z6(n) = \frac{3(1+n)(2+n)(3+n)(-9+\sqrt{33+4n(5+n)}+n(-7-2n+\sqrt{33+4n(5+n)}))}{(561+87\sqrt{33+4n(5+n)}+n(505+20n^2+45\sqrt{33+4n(5+n)}+6n(28+\sqrt{33+4n(5+n)})))^2}$$

For 1 TLM in the down state, non-resonance and a LMPN, $S_4$ is given by

$$S_4(\bar{n}, 1, \gamma t)_{NR} = \sum_{n=1}^{\infty} \frac{n\langle n|\rho_f(0)|n\rangle}{n+\Delta} \mathrm{Sin}^2\left(2\pi\sqrt{n+1+\Delta}\sqrt{n+\Delta}\,\tau\right) \quad (C5)$$

For 2 TLMs in the down state at resonance and a LMPN, $S_4$ is given by

$$S_4(\bar{n}, 2, \gamma t) = \langle 1|\rho_f(0)|1\rangle \mathrm{Sin}^2(2\sqrt{2}\pi\sqrt{n+1}\,\tau) + 8\sum_{n=2}^{\infty} \langle n|\rho_f(0)|n\rangle \left\{ \frac{n(n-1)}{(2n-1)^2} \mathrm{Sin}^2\left[2\pi\sqrt{n+1}\sqrt{n-\frac{1}{2}}\,\tau\right] \right.$$

$$\left. + \frac{n}{8(2n-1)^2} \mathrm{Sin}^2\left[4\pi\sqrt{n+1}\sqrt{n-\frac{1}{2}}\,\tau\right] \right\} \quad (C6)$$

### **APPENDIX D: APPROXIMATE SOLUTION FOR $S_1$**

Approximate solutions for $S_1$ and $S_4$ may be useful for reducing the calculation times needed for the exact expressions. However, it is necessary to consider the accuracy for using approximations. We start with approximations to the eigenvectors. Two good approximations were considered in [3]. They were called the Average Field Approximation (case 1) which was relevant when the mean number of photons is much greater than the number of TLMs which are initially in the up state and the Modified TLM approach (case 2) which was relevant for the mean number of photons being small and the system has big negative spin. Each approximation can be approached in two ways. The first is to make approximations to the Hamiltonian such as averaging over the field for case one and solve for the eigenvectors and eigenvalues. Unfortunately that techniques doesn't produce direct results. Instead for case one write the Hamiltonian as a sum over individual Hamiltonians, one for each TLM. For instance

$$H = \sum_{j=1}^{N} \frac{(a^\dagger a + R_3)}{N} - \kappa \sqrt{n_o} L R_{j+} - \kappa^* \sqrt{n_o} L^\dagger R_{j-} .  \tag{D1}$$

Recalling that c is a good quantum number, replace the first term by

$$c_j = \frac{c}{N} \tag{D2}$$

This may be interpreted as the Hamiltonian of N separate TLMs interacting with the electromagnetic field. Each interaction conserves the average energy c of the system but the TLMs "see" each other only through an average field. Therefore this problem can be treated by solving exactly for the eigenvalues and vectors for just one TLM, and the total state and energy of the system is formed from the single TLM in such a way as to give states of total r, c, and j, where j is an index representing the energy of the state. Note that $L$ is the photon lowering operator ($L \, |n> = |n-1>$) and $L^\dagger$ is the raising operator ($L^\dagger \, |n> = |n+1>$) and that $n_o$ is the mean number of photons.

The eigenvalues and eigenvectors for the ground and excited states, respectively, for this (single interacting) TLM case are easily found to be

$$\lambda_o = c_j - |\kappa|\sqrt{n_o} \tag{D3}$$

$$\overline{|\tfrac{1}{2}, c_j, 0>}_f = \frac{1}{\sqrt{2}} \left[ |c_j - \tfrac{1}{2}> |\tfrac{1}{2}, \tfrac{1}{2}> e^{i\varphi} + |c_j + \tfrac{1}{2}> |\tfrac{1}{2}, -\tfrac{1}{2}> \right] \tag{D4}$$

$$\lambda_1 = c_j + |\kappa|\sqrt{n_o} \tag{D5}$$

$$\overline{|\tfrac{1}{2}, c_j, 1>}_f = \frac{1}{\sqrt{2}} \left[ |c_j - \tfrac{1}{2}> |\tfrac{1}{2}, \tfrac{1}{2}> - |c_j + \tfrac{1}{2}> |\tfrac{1}{2}, -\tfrac{1}{2}> e^{-i\varphi} \right] \tag{D6}$$

These eigenvectors are generated from the non-interacting basis states by a rotation of 45° in the internal space of TLM plus field. The states $|\overline{r, c, j}>_f$ are constructed in a manner similar to that used by Dicke [38] and make full use of the definitions of the $|r, m>$ states. Construct a state from the single TLM states such that r and c are constants of motion and that this state has $n_+$ TLMs in the excited state and $n_-$ TLMs in the ground state. The total energy of the system is then

$$\begin{aligned}\lambda_j^{(f)} &= (n_+ + n_-)c_j + (n_+ - n_-)|\kappa|\sqrt{n_o} \\ &= c - 2|\kappa|\sqrt{n_o}(r - j)\end{aligned} \tag{D7}$$

The ground state is represented by j=0 and $n_-$=N=2r. The index m is not used here for $n_+ - n_-$ since m is reserved to represent the internal energy of the N non-interacting TLMs .

If the ground and excited TLM states are represented by $-$ and $+$, then the state $|\overline{r,c,j}>_f$ is constructed from these states in exactly the same way that the $|r,m>$ states are constructed from $|\uparrow>=|\frac{1}{2},\frac{1}{2}>$ and $|\downarrow>=|\frac{1}{2},-\frac{1}{2}>$ states. Details are provided in [3] for the resonant results

$$|\overline{r,c,j}>_f = \frac{1}{2^r\sqrt{\frac{(2r)!}{j!(2r-j)!}}} \sum_{L=0}^{2r} \sum_{L'=down}^{up'} e^{\frac{(L+L')\pi i}{2}}[L!(2r-L)!]$$

$$\div\left[\left(\frac{L+L'}{2}\right)!\left(\frac{L-L'}{2}\right)!\left(j-\left(\frac{L+L'}{2}\right)\right)!\left(n_- - \left(\frac{L-L'}{2}\right)\right)!\right] \quad (D8)$$

$$\times e^{(n_- - L)\varphi}\sqrt{\frac{(2r)!}{L!(2r-L)!}}|c-m>|r,m>,$$

where
$$n_- = 2r - j, \quad (D9)$$
$$m = r - L, \quad (D10)$$
$$down = max[-L, L - 2n_-], \quad (D11)$$
$$up = min[L, 2j - L]. \quad (D12)$$

and the prime on the second sum indicates that only every other value starting with "down" is used in the sum. Equation (D8) is the general $|\overline{r,c,j}>_f$ state expressed in terms of $|c-m>|r,m>$ states even for $r\neq N/2$.

For the non-resonant case the results are found in a similar manner in [24], namely;

$$|\frac{N}{2},c,j>_{AFA} = \frac{(b1)^j(a1)^{N-j}}{\sqrt{\frac{N!}{n_+!n_-!}}} \sum_{L=0}^{N} \sum_{L'=down}^{up} \left(\frac{b2a2}{b1a1}\right)^{\frac{L}{2}}\left(\frac{b2a1}{b1a2}\right)^{\frac{L'}{2}} e^{\frac{(L+L')\pi i}{2}}[L!(2r-L)!]$$

$$\div\left[\left(\frac{L+L'}{2}\right)!\left(\frac{L-L'}{2}\right)!\left(j-\left(\frac{L+L'}{2}\right)\right)!\left(n_- - \left(\frac{L-L'}{2}\right)\right)!\right]$$

$$\times e^{(N-j-L)\varphi}\sqrt{\frac{(2r)!}{L!(2r-L)!}}|c-m>|r,m>, \quad (D13)$$

where
$$n_+ = j$$
$$n_- = N - j,$$
$$m = \frac{N}{2} - L,$$
$$down = max[-L, L - 2n_-],$$
$$up = min[L, 2j - L].$$

The prime on the second sum indicates that only every other value starting with "down" is used in the sum. We already know that the eigenvectors can be expressed in the form ($r = \frac{N}{2}$ and $c$ is choosen as $n + \frac{N}{2}$)

$$\left|\frac{N}{2}, c, j\right> = \sum_{p=0}^{N} A_p^{\left(\frac{N}{2}, n+\frac{N}{2}, j\right)} |n+p\rangle \left|\frac{N}{2}, \frac{N}{2} - p\right>, \text{ when } n > r \tag{D14}$$

Making the correspondence to Eq. (D11), we can perform the substitution p=L and

$$down = \begin{matrix} -p & j < N-p \\ p+2j-2N & j \geq N-p \end{matrix}$$

$$up = \begin{matrix} p & j \geq p \\ 2j-p & j < p \end{matrix} \tag{D15}$$

express $A_{n+p}^{(r,c,j)}$ after some algebraic manipulation as

$$A_{n+p}^{\left(\frac{N}{2}, n+\frac{N}{2}, j\right)} = (b1)^j (a1)^{N-j} \begin{cases} \left(\frac{a2}{a1}\right)^p \sqrt{\frac{(N-j)!(N-p)!}{p!j!}} \frac{F\left(-j, -p, N+1-p-j; -\left(\frac{b2a1}{b1a2}\right)\right)}{(N-j-p)!} & N \geq j+p \\ (-1)^{p-N} \left(\frac{b2a2}{b1a1}\right)^p \left(\frac{b2a1}{b1a2}\right)^{-N} \sqrt{\frac{p!j!}{(N-p)!(N-j)!}} \frac{F\left(j-N, p-N, 1+p+j-N; -\left(\frac{b2a1}{b1a2}\right)\right)}{(j+p-N)!} & N < j+p \end{cases} \tag{D16}$$

where F is the hypergeometric function. The quantities (a1, a2, b1, b2) are the coefficients for the exact solution for the single TLM in the AFA and are given by

$$a1 = \frac{1}{\sqrt{1 + [0.5\beta - 0.5\sqrt{4.+\beta^2}]^2}}$$

$$a2 = \frac{-0.5\beta + 0.5\sqrt{4.+\beta^2}}{\sqrt{1 + [0.5\beta - 0.5\sqrt{4.+\beta^2}]^2}}$$

$$b1 = \frac{1}{\sqrt{1 + [0.5\beta + 0.5\sqrt{4.+\beta^2}]^2}} \tag{D17}$$

$$b2 = \frac{0.5\beta + 0.5\sqrt{4.+\beta^2}}{\sqrt{1 + [0.5\beta + 0.5\sqrt{4.+\beta^2}]^2}}$$

Further the effective eigenvalues in the AFA lie along a straight line and are given by

$$q = -\left(\frac{N}{2} + n\right)\bar{\beta} + (r-j)\sqrt{4+\bar{\beta}^2} \tag{D18}$$

and

$$\bar{\beta} = \frac{\omega - \Omega}{|\kappa|\Omega\sqrt{n_o}} \tag{D19}$$

In order to find the approximate value for $S_1$ it is necessary to insert the values for q and $A_{n+p}^{(r,c,j)}$ into Eq. (B11). Insertion of q is straightforward; however, determining

$$(A^*)_n^{\frac{N}{2},(n+\frac{N}{2}),j} A_n^{\frac{N}{2},(n+\frac{N}{2}),j'} \sum_{p=0}^{N} p \, A_{n+p}^{\frac{N}{2},(n+\frac{N}{2}),j} (A^*)_{n+p}^{\frac{N}{2},(n+\frac{N}{2}),j'} \tag{D20}$$

requires some thought. In particular the summation over p can be difficult to determine but it was found from near first principal that this summation can be found to exist for $j' = j + 1$ and is very small for other values of $j'$. The details are found in [24 Appendix B].

Once Eq. (D20) is found, the approximate form for $S_1(\widetilde{\bar{n},N},\gamma t)_{AFA}$ can be written as

$$S_1(\widetilde{\bar{n},N},\gamma t)_{AFA} = N \sum_{n=0}^{\infty} \langle n|\rho_f(0)|n\rangle \frac{n_0}{n_0 + \Delta} Sin^2\left[\sqrt{n_0 + \Delta}\Omega|\kappa|t\right]. \tag{D21}$$

Again refer to ref. [24] for the details. Note that this result may not exactly match the exact results which was found to be the case for $S_4$ which is discussed below. We will come back to a better results later.

### Appendix E; Modified TLM Approximation

As in the case for stimulated emission, it is possible to obtain an approximate solution for stimulated absorption by making use of the approximate eigenvectors and eigenvalues seen in reference [3]. However, we will generate the solution for the non-resonant case which was not done in the earlier work. The approximation used is based on a Modified TLM approach which is valid when the number of photons is much less than the number of TLMs and the TLMs are essentially in the ground state.

Starting with Eq. (A13), the exact solution for the $A_n^{(r,c,j)}$ is found by unraveling from one end with the details described in ref. [3]. Since we will find the solution for the non-resonant case as we did for the resonant case for the modified TLM approximation in [3] in terms of the states with just 1 photon and N TLMs all starting in the ground state we need the solutions for (A13) for c = 1-r. Equation (A13) for this case then becomes

$$\begin{aligned}&-|\kappa|e^{-i\varphi}\sqrt{n}C_{r,1-r-n}A_{n-1}^{(r,1-r,j)} + \left(c + \frac{\omega - \Omega}{\Omega}n - \lambda_{r,1-r,j}\right)A_n^{(r,1-r,j)} \\ &- |\kappa|e^{i\varphi}\sqrt{n+1}C_{r,-r-n}A_{n+1}^{(r,1-r,j)} = 0\end{aligned} \tag{E1}$$

There will only be 2 components for the eigenvectors for this case, namely $A_0^{(r,1-r,j)}$ and $A_1^{(r,1-r,j)}$. Start by setting $A_0^{(r,1-r,j)}=1$ although that is the un-normalized value. Set n=0 in (E1) and solve for $A_1^{(r,1-r,j)}$. The results is

$$A_1^{(r,1-r,j)} = \frac{(c - \lambda_{r,1-r,j})}{|\kappa|e^{i\varphi}\sqrt{N}}. \tag{E2}$$

Further, the eigenvalues can then be found by applying eq. (E1) and using $\tilde{A}_2^{(r,c,j)} = 0$, thus

$$-|\kappa|e^{-i\varphi}\sqrt{N} + \frac{(c - \lambda_{r,c,j})\left(c + \frac{\omega - \Omega}{\Omega}n - \lambda_{r,c,j}\right)}{|\kappa|e^{i\varphi}\sqrt{N}} = 0. \tag{E3}$$

The solution for the $q$ is given for (c=1-r or one photon) is

$$|\kappa|q_{r,1-r,j} = \frac{-\beta \pm \sqrt{\beta^2 + 4|\kappa|^2 N}}{2} = |\kappa|\left[-\frac{\bar{\beta}}{2} \pm \sqrt{\Delta + N}\right] = c - \lambda_{r,1-r,j} \tag{E4}$$

In this equation, $\beta = \frac{\omega - \Omega}{\Omega}, \bar{\beta} = \frac{\omega - \Omega}{|\kappa|\Omega}$ and $\Delta = \frac{\bar{\beta}^2}{4}$.

With this, using proper normalization and after a little work, the approximate ground and excited states for one photon and initial all down TLMs are given as follows;

$$
\begin{aligned}
|r, 1-r, 0>_{NR} &= \left[d_1|0> |r, -r+1> + e^{-i\varphi} d_2 |1> |r, -r>\right] \\
|r, 1-r, 1>_{NR} &= \left[e_1|0> |r, -r+1> - e^{-i\varphi} e_2 |1> |r, -r>\right] \\
d_1 &= N_0, \quad d_2 = N_0 \frac{\sqrt{\Delta} + \sqrt{\Delta + N}}{\sqrt{N}}, \quad e_1 = N_1, \\
e_2 &= N_1 \frac{\sqrt{\Delta + N} - \sqrt{\Delta}}{\sqrt{N}} \\
N_0 &= \sqrt{\frac{N}{2(N + \Delta) - \beta\sqrt{N + \Delta}}} \\
N_1 &= \sqrt{\frac{N}{2(N + \Delta) + \beta\sqrt{N + \Delta}}}
\end{aligned}
\tag{E5}
$$

For higher numbers of photons Eq. (A13) could be further unraveled. Instead we follow the solution suggested in references [3 and 24].

The potential eigenvalues for n photons rather than n=1 is given by

$$\tilde{q}_{r,c,j} = \left[-\frac{\bar{\beta}n}{2} + (n-2j)\sqrt{\Delta + N}\right] = c - \tilde{\lambda}_{r,c,j} \tag{E6}$$

The eigenstate is provided by the product of the number of up states (j=1) and the down states (j=0) and is given by inspection as in reference [3] to be

$$
|\widetilde{r, n, j}>_{NR} = \frac{1}{\sqrt{\frac{n!}{n_+! n_-!}}} \sum_\rho \left\{ \prod_{k=1}^{n_+} \left([e_1|0> |r, -r+1>] + e_2|1> |r, -r> e^{i(\pi-\varphi)}\right)_k \right. \\
\left. \times \prod_{k'=1}^{n_-} \left(d_1|0> |r, -r+1> + d_2 e^{-i\varphi}|1> |r, -r>\right)_{k'} \right\},
\tag{E7}
$$

where $\rho$ represents the various permutations on the ground and excited states and k and $k'$ represent the different excited and ground state TLMs interacting with the field. This state may be represented by a double sum

$$
|\widetilde{r, n, j}>_{NR} = \frac{1}{\sqrt{\frac{n!}{n_+! n_-!}}} \sum_{k=0}^{n_+} \sum_{k'=0}^{n_-} (e1)^{n_+ - k} (e2)^k (d1)^{n_- - k'} \\
\times (d2)^{k'} e^{ik(\pi-\varphi)} e^{i(-k')\varphi} \left[\frac{n_+!}{k!(n_+ - k)!}\right]\left[\frac{n_-!}{k'!(n_- - k')!}\right]\left[\frac{n!}{n_+! n_-!}\right] \div \left[\frac{n!}{(k+k')!(n-k-k')!}\right] \\
\times \sqrt{\frac{n!}{(k+k')!(n-k-k')!}} |k+k'> |r, -r+n-(k+k')>
\tag{E8}
$$

This is an unusual construction since the product affects the number of photons but not the number of TLMs. The first phase factor and first factorial term comes from the phase and number of terms encountered in the product over the individual excited states. The second phase factor and second factorial term comes from the product over individual ground states. The third factorial term comes from the number of permutations over the $n_+$ (excited) and $n_-$ (ground) states. The fourth factorial term comes from the number of terms necessary to form the basis states and the square root term is necessary so that the basis states are normalized. Terms can be combined to simplify the expression; namely

$$|\widetilde{r,n,j}>_{NR} = \sum_{k=0}^{n_+} \sum_{k'=0}^{n_-} \frac{\sqrt{n_+! n_-! (k+k')! (n-(k+k'))!}}{k! (n_+ - k)! k'! (n_- - k')!} \times (e1)^{n_+ - k} (e2)^k (d1)^{n_- - k'} (d2)^{k'} e^{ik(\pi - \varphi)} e^{i(-k')\varphi} \ |k+k'>|r, -r+n-(k+k')> \quad (E9)$$

Change variables letting $L = k + k', L' = k - k', n_+ = j, n_- = n - j$ and with some thought

$$|\widetilde{r,n,j}>_{NR} = (e1)^j (d1)^{n-j} \sum_{L=0}^{N} \sum_{L'=Max[-L, L=2j-2n]}^{Min[L, 2j-L]} \sqrt{j!(n-j)! L! (n-L)!}$$

$$\div \left[ \left(\frac{L+L'}{2}\right)! \left(j - \frac{L+L'}{2}\right)! \left(\frac{L-L'}{2}\right)! \left(n - j - \left(\frac{L-L'}{2}\right)\right)! \right] \quad (E10)$$

$$\times (e1)^{-\left(\frac{L+L'}{2}\right)} (e2)^{\left(\frac{L+L'}{2}\right)} (d1)^{-\left(\frac{L-L'}{2}\right)} (d2)^{\left(\frac{L-L'}{2}\right)} e^{i\frac{L+L'}{2}(\pi - \varphi)} e^{-i\left(\frac{L-L'}{2}\right)\varphi} |L>|r, -r+n-L>$$

Where the prime on the L' indicates that only every other term in the summation over L' is taken. Using the correspondence between equation (A12) and (E10),

$$\tilde{A}_L^{r,c,j} = (e1)^j (d1)^{n-j} \left(\frac{d2e2}{e1d1}\right)^L \sqrt{j!(n-j)! L! (n-L)!} \, e^{i\frac{L}{2}(\pi - 2\varphi)}$$

$$\times \sum_{L'=Max[-L, L=2j-2n]}^{Min[L, 2j-l]} \left(\frac{d1e2}{e1d2}\right)^{L'} e^{i\frac{L'}{2}\pi} \div \left[ \left(\frac{L+L'}{2}\right)! \left(j - \frac{L+L'}{2}\right)! \left(\frac{L-L'}{2}\right)! \left(n - j - \left(\frac{L-L'}{2}\right)\right)! \right] \quad (E11)$$

There are 2 separate expressions depending on the lower limit of the summation. In one $L \leq n - j$ while for the other $L > n - j$. For both cases a change in variable and the fact that the summation is over every other term results in

$$A_L^{r,n-r,j} = (e1)^j (d1)^{n-j} \left(\frac{d2}{d1}\right)^L e^{-iL\varphi} \sqrt{j!(n-j)!} \sqrt{L!(n-L)!}$$

$$\times \sum_{jj=0}^{Min[L,j]} \left(-\frac{d1e2}{d2e1}\right)^{jj} \div [(jj)! (L-jj)! (j-jj)! (n-j-L+jj)!] \quad L \leq n - j \quad (E12)$$

$$A_L^{r,n-r,j} = \sqrt{j!(n-j)!} (e1)^j (d1)^{n-j} \left(\frac{e2}{e1}\right)^L \left(\frac{d1e2}{d2e1}\right)^{\frac{2j-2n}{2}} e^{i(j-n)\pi} e^{iL\pi} e^{-iL\varphi} \sqrt{L!(n-L)!}$$

$$\times \sum_{jj}^{Min[n-j, n-L]} \left(-\frac{d1e2}{d2e1}\right)^{jj} \div [(jj+L+j-n)! (n-j-jj)! (n-L-jj)! (jj)!] \quad L > n - j \quad (E13)$$

By using the definition of the hypergeometric function, equations (E12 and E13) become

$$A_L^{r,n-r,j} = (e1)^j (d1)^{n-j} \left(\frac{d2}{d1}\right)^L e^{-iL\varphi} \sqrt{\frac{(n-j)!(n-L)!}{j!L!}}$$
$$\times \frac{F\left(-L,-j,n-j-L+1;-\frac{d1e2}{d2e1}\right)}{(n-j-L)!} \text{ when } L \leq n-j \tag{E14}$$

and

$$A_L^{r,n-r,j} = (e1)^j (d1)^{n-j} \left(\frac{e2}{e1}\right)^L \left(\frac{d1e2}{d2e1}\right)^{j-n} e^{i(j-n)\pi} e^{iL\pi} e^{-iL\varphi} \sqrt{\frac{L!j}{(n-j)!(n-L)!}}$$
$$\times \frac{F\left(j-n, L-n, j+L-n+1; -\frac{d1e2}{d2e1}\right)}{(j+L-n)!} \quad L > n-j \tag{E15}$$

With these expressions, it is not possible to determine the approximate solutions for $S_4(\bar{n}, N, \gamma t)$ directly. First it is necessary to evaluate the summation over p which is found in Eq. (B13), that is

$$\sum_{p=0}^{\min[N,n]} pA_{n-p}^{\frac{N}{2}, n-\frac{N}{2}, j} (A^*)_{n-p}^{\frac{N}{2}, n-\frac{N}{2}, j'}.$$

In reference (24 Appendix B), it was demonstrated that if the eigenvalues of the Hamiltonian lie along a straight line and the appropriate changes between the stimulated emission and stimulated absorption are made then

$$\sum_{p=0}^{n} pA_{n-p}^{r,n-r,j} (A^*)_{n-p}^{r,n-r,j'} = \frac{\sqrt{j+1}\sqrt{n-j}\delta_{j',j+1}}{2\sqrt{1+\frac{\Delta}{N}}} \text{ when } j' \neq j. \tag{E16}$$

This is the same argument made to simplify Eq. (D20). Further

$$A_n^{r,n-r,j} = (e2)^j (d2)^{n-j} e^{ij\pi} e^{-in\varphi} \sqrt{\frac{n!}{j!(n-j)!}}. \tag{E17}$$

Finally

$$(A^*)_n^{\frac{N}{2},n-\frac{N}{2},j} A_n^{\frac{N}{2},n-\frac{N}{2},j'} \sum_{p=0}^{\min[N,n]} p A_{n-p}^{\frac{N}{2},n-\frac{N}{2},j} (A^*)_{n-p}^{\frac{N}{2},n-\frac{N}{2},j'}$$

$$\cong -(e2)^{2j+1}(d2)^{2n-2j-1} \frac{n!}{\sqrt{j!(n-j)!(j+1)!(n-j-1)!}} \frac{\sqrt{j+1}\sqrt{n-j}\,\delta_{j',j+1}}{2\sqrt{1+\frac{\Delta}{N}}}$$

$$= -(e2)^{2j+1}(d2)^{2n-2j-1} \frac{n}{2\sqrt{1+\frac{\Delta}{N}}} \frac{(n-1)!}{j!(n-j-1)!} \tag{E18}$$

$$= -\frac{n(d2e2)}{2\sqrt{1+\frac{\Delta}{N}}} \frac{(n-1)!\,[(e2)^2]^j[(d2)^2]^{n-j-1}}{j!(n-j-1)!}$$

$$= -\frac{n}{4} \frac{N}{N+\Delta} \frac{(n-1)!\,[(e2)^2]^j[(d2)^2]^{n-j-1}}{j!(n-j-1)!}.$$

The only restriction on equation (E18) is that n<N. With the use of equations (E5, E6 and E18) and a considerable amount of algebra we find

$$S_4(\widetilde{\bar{n},N},\gamma t) \cong \sum_{n=0}^{\infty} \langle n|\rho_f(0)|n\rangle n \frac{N}{N+\Delta} Sin^2\{\sqrt{N+\Delta}\Omega|\kappa|t\}, N \gg \bar{n} \tag{E19}$$

The "goodness" of the approximate solution (Eq. (E19)) depends on the approximations made above. This can only be examined by performing the exact calculations for $S_4$ and comparing to $\widetilde{S_4}$.

This comparison shows that a slightly better solution for the approximated value of $S_4(\bar{n},N,\gamma t)$ and $\tilde{q}_{r,c,j}$ are given by

$$S_4(\widetilde{\bar{n},N},\gamma t) \cong \sum_{n=0}^{\infty} \langle n|\rho_f(0)|n\rangle n \frac{N-\frac{n}{2}}{N-\frac{n}{2}+\Delta} Sin^2\left\{\sqrt{N-\frac{\bar{n}}{2}+\Delta}\Omega|\kappa|t\right\}$$

$$\tilde{q}_{r,c,j} = \left[-\frac{\bar{\beta}n}{2} + (n-2j)\sqrt{\Delta+N-\frac{\bar{n}}{2}}\right] N \gg \bar{n} \tag{E20}$$

Unfortunately Eq. (E20) still does not give appear to give good results st time becomes large. The location of echoes is good but the amplitude does not decrease appropriately. This is demonstrated in the Appendix F.

### APPENDIX F: FURTHER DISCUSSION ON THE GOODNESS OF THE APPROXIMATE RESULTS

Since calculations using the approximate solution are much faster that for the exact results, there is an interest in modifying the approximate results to match the exact results. So shifting focus back to the approximate expression for $S_4$ given by Eq. (E20), we perform the calculation for the exact and approximate results to greater times where it is even more apparent that there is a significant difference between the exact and approximate results. Again using the coherent distribution with a mean photon number of 10 the results for time are calculated out to 100 $\tau_R$ is shown in Figure F1.

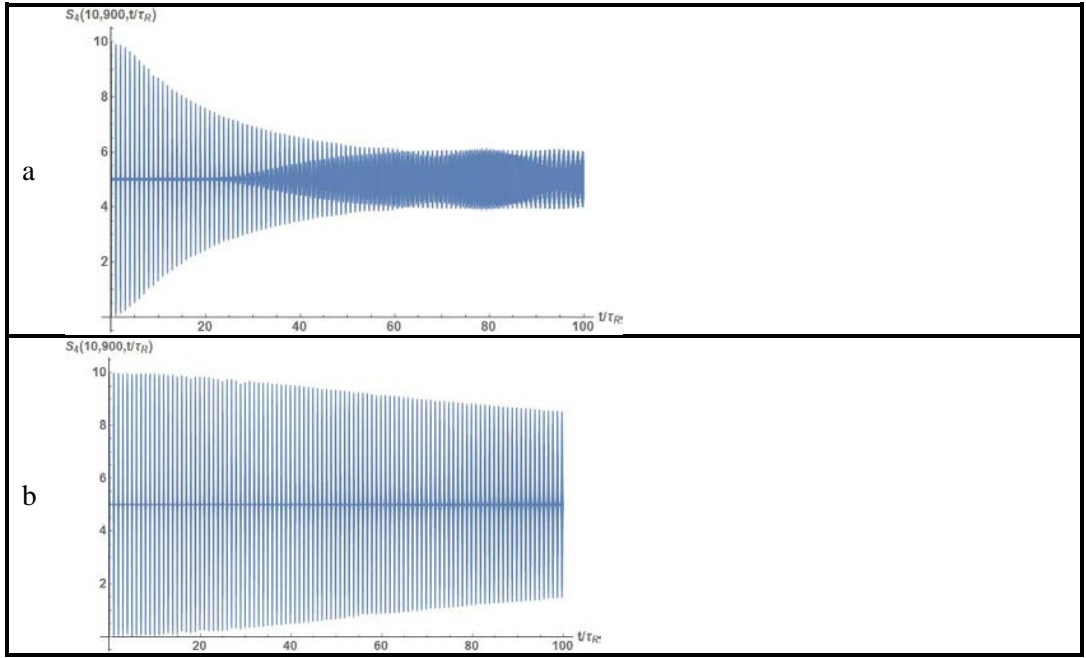

*Figure F1: Exact (a) and approximate (b) results for stimulated absorption for 900 TLMs and a mean photon number of 10 for the coherent photon distribution out to a normalized time of 100.*

Further examination of the exact solution is necessary to consider each of the factors used in the approximate solution. Towards that end we first consider Eq. (E16) to determine its accuracy. Examination of the exact solution reveals that the condition that j'=j±1 is fairly good. Terms for j'=j±2 are 50 times smaller than for j±1. Terms for j' larger than j+2 are many orders of magnitude smaller and can be completely ignored. In fact for j'=j+1 Eq. (E16) is nearly exact to within a percent. Thus without much loss of accuracy Eq. (B13) can be approximated by

$$S_4(\bar{n}, N, \gamma t) \cong -4 \sum_{n=0}^{\infty} \langle n|\rho_f(0)|n\rangle \sum_{j=0}^{Min[N,n]-1} Sin^2\left\{\frac{\left[q_{\frac{N}{2},(n-\frac{N}{2}),j} - q_{\frac{N}{2},(n-\frac{N}{2}),j+1}\right]}{2} \Omega|\kappa|t\right\}$$
$$\times (A^*)_n^{\frac{N}{2},n-\frac{N}{2},j} A_n^{\frac{N}{2},n-\frac{N}{2},j+1} \frac{\sqrt{j+1}\sqrt{n-j}}{2\sqrt{1+\frac{\Delta}{N}}}$$
(F1)

A further examination of the difference between the exact and approximate solution (5 and 6), leads to question the accuracy of Eqs. (E14 and E15) compared to the exact solutions for the eigenvectors. It is found that as n increases there is an increasing difference between the exact and approximate results. To illustrate that difference a table containing results for n=10 and n=25 for the values of $(A^*)_n^{\frac{N}{2},n-\frac{N}{2},j} A_n^{\frac{N}{2},n-\frac{N}{2},j+1}$ for the resonant case with the minus sign neglected

Table D1: Exact and Approximate values for $(A^*)_n^{\frac{N}{2},n-\frac{N}{2},j} A_n^{\frac{N}{2},n-\frac{N}{2},j+1}$

| Exact $(A^*)_{10}^{\frac{N}{2},10-\frac{N}{2},j} A_{10}^{\frac{N}{2},10-\frac{N}{2},j+1}$ | Approximate. $(A^*)_{10}^{\frac{N}{2},10-\frac{N}{2},j} A_{10}^{\frac{N}{2},10-\frac{N}{2},j+1}$ | Exact $(A^*)_{25}^{\frac{N}{2},25-\frac{N}{2},j} A_{25}^{\frac{N}{2},25-\frac{N}{2},j+1}$ | Approximate. $(A^*)_{25}^{\frac{N}{2},25-\frac{N}{2},j} A_{25}^{\frac{N}{2},25-\frac{N}{2},j+1}$ |
|---|---|---|---|
| .00315 | 0.003088 | 1.738E-07 | 1.49E-07 |
| .02095 | 0.020716 | 2.934E-06 | 2.58E-06 |
| .07208 | 0.071762 | 2.75E-05 | 2.48E-05 |
| .15502 | 0.155024 | .0001748 | 0.000161 |
| .2242 | 0.224652 | .000824 | 0.000773 |
| .2242 | 0.224652 | .003033 | 0.002891 |
| .15502 | 0.155024 | .00899 | 0.008696 |
| .07208 | 0.071762 | .02195 | 0.021489 |
| .02095 | 0.020716 | .0448 | 0.044301 |
| .00315 | 0.003088 | .07727 | 0.077015 |
| | | .1135 | 0.113758 |
| | | .1427 | 0.143485 |
| | | .1539 | 0.154981 |
| | | .1427 | 0.143485 |
| | | .1135 | 0.113758 |
| | | .07727 | 0.077015 |
| | | .0448 | 0.044301 |
| | | .02195 | 0.021489 |
| | | .00899 | 0.008696 |
| | | .003033 | 0.002891 |
| | | .000824 | 0.000773 |
| | | .0001748 | 0.000161 |
| | | 2.75E-05 | 2.48E-05 |
| | | 2.934E-06 | 2.58E-06 |
| | | 1.738E-07 | 1.49E-07 |

As can be seen the exact values are slightly larger at the ends while the approximate values are slightly larger in the middle. If one were to ignore the j dependence within the sin function, the resulting summation over j would result in a value of $\frac{n}{4}$ for both the exact and approximate values. This implies that the source of the differences between the exact and approximate solution to $S_4$ are in the j variation of the eigenvalues. The approximate eigenvalues results in the use of $\sqrt{N - \frac{n}{2} + \Delta}$ in the approximate solution for $S_4$. The exact expression for $\frac{\left[q_{\frac{N}{2},(n-\frac{N}{2}),j} - q_{\frac{N}{2},(n-\frac{N}{2}),j+1}\right]}{2}$ is slightly dependent on j and the dependence on n does not match the approximate expression given above. For example, assume the resonant case for $\Delta = 0$. In that case the exact results is much closer to $\sqrt{N - \frac{n-1}{2}}$ with a variation in the second decimal place. For normalized time larger than 10 this is enough to generate large differences in the $Sin^2$ function which is enough to cause the differences in the amplitude decay in the revivals seen between the approximate and the exact result. This means that a more exact approximate results for $S_4$ is given by

$$S_4(\widetilde{\bar{n},N},\gamma t) \cong \sum_{n=0}^{\infty} n \frac{2N - n + 1}{2N + 2\Delta - n + 1} \langle n|\rho_f(0)|n\rangle$$
$$\times \sum_{j=0}^{Min[N,n]-1} Sin^2\left\{\frac{\left[q_{\frac{N}{2},(n-\frac{N}{2}),j} - q_{\frac{N}{2},(n-\frac{N}{2}),j+1}\right]}{2}\Omega|\kappa|t\right\} \frac{(n-1)![(e2)^2]^j[(d2)^2]^{n-j-1}}{j!(n-j-1)!}, \quad (F2)$$

We have made a minor additional change by replacing N by $N - \frac{n-1}{2}$ in Eq. (F2)

As I mentioned earlier it is probably necessary to modify Eq. (D21) and instead of using the approximate eigenvalues it is necessary to use the exact results. If one were to do that Eq. (D21) would become using normalized time

$$S_1(\widetilde{\bar{n},N},\gamma t) = \frac{4N(N-1)!}{2\sqrt{2+\beta}} \sum_{n=0}^{\infty} \langle n|\rho_f(0)|n\rangle \sum_{j=1}^{N-1} \frac{[(b1)^2]^j[(a1)^2]^{N-j-1}}{j!(N-1-j)!\sqrt{1+\Delta}}$$
$$\times Sin^2\left\{\pi\sqrt{\bar{n}+1+\Delta}\left[q_{\frac{N}{2},(n+\frac{N}{2}),j} - q_{\frac{N}{2},(n+\frac{N}{2}),j+1}\right]\tau\right\} \quad (F5)$$

For the resonant case, $S_1$ and $S_4$ become

$$S_1(\widetilde{\bar{n},N},\gamma t)_R = \frac{4N!}{2^N\sqrt{2}} \sum_{n=0}^{\infty} \langle n|\rho_f(0)|n\rangle \sum_{j=1}^{N-1} \frac{Sin^2\left\{\pi\sqrt{\bar{n}+1+\Delta}\left[q_{\frac{N}{2},(n+\frac{N}{2}),j} - q_{\frac{N}{2},(n+\frac{N}{2}),j+1}\right]\tau\right\}}{j!(N-1-j)!} \quad (F4)$$

$$S_4(\widetilde{\bar{n},N},\gamma t) \cong \sum_{n=0}^{\infty} \frac{n!}{2^{n-1}} \langle n|\rho_f(0)|n\rangle \sum_{j=0}^{Min[N,n]-1,} \frac{Sin^2\left\{2\pi\sqrt{N+1}\left[q_{\frac{N}{2},(n-\frac{N}{2}),j} - q_{\frac{N}{2},(n-\frac{N}{2}),j+1}\right]\tau\right\}}{j!(n-j-1)!}. \quad (F5)$$

### Appendix G: Entropy

Entropy may be used to indicate the presence of entanglement or dis-entanglement. In general, the authors [12, 13, 29, 32-35] relate entanglement to the Von Neumann entropy defined as

$$S(\rho) = Tr(\rho Ln\rho), \quad (G1)$$

with $\rho$ being the density matrix for the system of TLMs and photons. However, for a closed system, this quantity should be either zero or a constant. Jarvis, et al [29] stated that for such a system the quantity of interest is the entropy of the reduced density matrix. This is found by taking the trace of $\rho$ over the field or over the TLMs. The resulting entropies are identical [12, 13], namely:

$$S(\rho_{TLM}) = Tr\rho_{TLM} Ln\rho_{TLM} = S(\rho_f) = Tr\rho_f Ln\rho_f. \tag{G2}$$

This equation is difficult to evaluate since the expansion of trace involves elements which are the expectation values of $\rho_f$ or $Ln\rho_f$. Between different values on n such as $\langle n|\rho_f|n'\rangle$. To simplify evaluation, the Shannon TLM reduced entropy [12, 13] is defined as

$$S_s(\rho_f) = Tr\rho_f Ln\rho_f = \sum_{n=0}^{\infty} \langle n|\rho_f|n\rangle Ln\langle n|\rho_f|n\rangle. \tag{G3}$$

There are again to cases of interest. The first is that there are few TLMs all in the up state and a LMPN with the second being a large number of TLMs all in the down state and a SMPN. Using techniques similar to finding the field $\langle E^- E^+(t)\rangle$:

$$\langle n|\rho_f(t)|n\rangle = \sum_{p=0}^{N} \langle n+p|\rho_f(0)|n+p\rangle \left[ \left( \sum_{j=0}^{Min[N,n+p]} (A^*)_{n+p}^{\frac{N}{2},n-\frac{N}{2}+p,j} A_n^{\frac{N}{2},n-\frac{N}{2}+p,j} Cos\left(\lambda_{\frac{N}{2},n-\frac{N}{2}+p,j} t\right) \right)^2 \right. $$
$$\left. + \left( \sum_{j=0}^{Min[N,n+p]} (A^*)_{n+p}^{\frac{N}{2},n-\frac{N}{2}+p,j} A_n^{\frac{N}{2},n-\frac{N}{2}+p,j} Sin\left(\lambda_{\frac{N}{2},n-\frac{N}{2}+p,j} t\right) \right)^2 \right] \tag{G4}$$

Insertion of Eq. (G4) into (G3) yields

$$S_s(\rho_f) = \sum_{n=0}^{\infty} \sum_{p=0}^{N} \langle n+p|\rho_f(0)|n+p\rangle G(n+p,n) Ln\left[ \sum_{p=0}^{N} \langle n+p|\rho_f(0)|n+p\rangle G(n+p,n) \right]. \tag{G5}$$

Let n+p = z, then

$$S_s(\rho_f) = \sum_{n=0}^{\infty} \sum_{z=n}^{N+n} \langle z|\rho_f(0)|z\rangle G(z,n) Ln\left[ \sum_{z=n}^{N+n} \langle z|\rho_f(0)|z\rangle G(z,n) \right], \tag{G6}$$

where

$$G(z,n) = \left[ \left( \sum_{j=0}^{Min[N,z]} (A^*)_z^{\frac{N}{2},z-\frac{N}{2},j} A_n^{\frac{N}{2},z-\frac{N}{2},j} Cos\left(\lambda_{\frac{N}{2},z-\frac{N}{2},j} t\right) \right)^2 + \left( \sum_{j=0}^{Min[N,z]} (A^*)_z^{\frac{N}{2},z-\frac{N}{2},j} A_n^{\frac{N}{2},z-\frac{N}{2},j} Sin\left(\lambda_{\frac{N}{2},z-\frac{N}{2},j} t\right) \right)^2 \right]. \tag{G7}$$

For the resonant case, only one term in Eq. (G7) is non-zero depending on whether z + n is even or odd.

$$G(z,n) = \left[\left(\sum_{j=0}^{Floor[\frac{Min[N,z]}{2}]} (A^*)_z^{\frac{N}{2},z-\frac{N}{2},j} A_n^{\frac{N}{2},z-\frac{N}{2},j} Cos\left(\lambda_{\frac{N}{2},z-\frac{N}{2},j}t\right)\right)^2\right], \quad z+n \text{ even} \tag{G8}$$

$$G(z,n) = \left[\left(\sum_{j=0}^{Floor[\frac{Min[N,z]}{2}]} (A^*)_z^{\frac{N}{2},z-\frac{N}{2},j} A_n^{\frac{N}{2},z-\frac{N}{2},j} Sin\left(\lambda_{\frac{N}{2},z-\frac{N}{2},j}t\right)\right)^2\right], \quad z+n \text{ odd} \tag{G9}$$

This is the results for stimulated absorption. It is possible to obtain results for stimulated emission as well. In order to do so, the value of $\langle n|\rho_f(t)|n\rangle$ must be found for the TLMs in the up state. Namely

$$\langle n|\rho_f(t)|n\rangle = \sum_{p=0}^{N} \langle n-p|\rho_f(0)|n-p\rangle \left|\sum_{j=0}^{N} (A^*)_{n-p}^{r,n+r-p,j} A_n^{r,n+r-p,j} e^{-i\lambda_{r,n+r-p,j}t}\right|^2 \tag{G10}$$

Equation (G10) is similar to Eq. (G4) and can in fact be written the same way. Note that the $\lambda$ can be replaced by the q as was the above case when performing calculations. Thus for the TLMs initially all up and n>>N

$$S_s(\rho_f)_{TLMs\ all\ up} = \sum_{n=0}^{\infty}\sum_{p=0}^{N} \langle n-p|\rho_f(0)|n-p\rangle GG(n-p,n) Ln\left[\sum_{p=0}^{N} \langle n-p|\rho_f(0)|n-p\rangle GG(n-p,n)\right]. \tag{G11}$$

$$GG(z,n) = \left[\left(\sum_{j=0}^{N}(A^*)_z^{\frac{N}{2},z+\frac{N}{2},j} A_n^{\frac{N}{2},z+\frac{N}{2},j} Cos\left(q_{\frac{N}{2},z-\frac{N}{2},j}|\kappa|t\right)\right)^2 + \left(\sum_{j=0}^{N}(A^*)_z^{\frac{N}{2},z+\frac{N}{2},j} A_n^{\frac{N}{2},z+\frac{N}{2},j} Sin\left(q_{\frac{N}{2},z-\frac{N}{2},j}|\kappa|t\right)\right)^2\right]. \tag{G12}$$

**Appendix H: The Q function**

The Q function is defined in [36] be a measure of the time behavior of the coherent state as a function of time. This function, $Q(\alpha,t)$, for big negative spin (BNS) and a SMPN photon distribution is define as the expectation value of the field density matrix in terms of the coherent state $|\alpha>$.

$$Q(\alpha,t) = \langle\alpha|\rho_f(t)|\alpha\rangle = e^{-|\alpha|^2}\sum_{s=0}^{\infty}\sum_{r=0}^{\infty}\frac{|\alpha|^s|\alpha|^r}{\sqrt{s!r!}}e^{-i\varphi(r-s)} <s|\rho_f(t)|r>. \tag{H1}$$

The density matrix of the field as a function of time, assuming an initial coherent field (This is by far the easiest to evaluate), is given by

$$< s|\rho_f(t)|r> = \sum_{p=0}^{N} \sum_{j=0}^{Min[N,r+p]} \sum_{j'=0}^{Min[N,s+p]} (A^*)_{r+p}^{\frac{N}{2},r-\frac{N}{2}+p,j} A_r^{\frac{N}{2},r-\frac{N}{2}+p,j} e^{-i\lambda_{\frac{N}{2},r-\frac{N}{2}+p,j}t} \tag{H2}$$
$$\times (A^*)_s^{\frac{N}{2},s-\frac{N}{2}+p,j'} A_{s+p}^{\frac{N}{2},s-\frac{N}{2}+p,j'} e^{iq_{\frac{N}{2},s-\frac{N}{2}+p,j'}t} \langle r+p|\rho_f(0)|s+p\rangle.$$

With the assumption that all the TLMs are initially in the down state and the initial photon distribution is the coherent state, Eq. (H1) with the mean photon number = $\bar{n}$, $Q(\alpha,t)$ can be written using Eq. (H2) as

$$Q(\alpha,t) = e^{-|\alpha|^2 - \bar{n}} \sum_{p=0}^{N} \left| \sum_{s=0}^{\infty} \frac{|\alpha|^s \bar{n}^{\frac{s+p}{2}} e^{is\varphi}}{\sqrt{s!(s+p)!}} \sum_{j'=0}^{Min[N,s+p]} (A^*)_s^{\frac{N}{2},s-\frac{N}{2}+p,j'} A_{s+p}^{\frac{N}{2},s-\frac{N}{2}+p,j'} e^{-iq_{\frac{N}{2},s-\frac{N}{2}+p,j'}t} \right|^2. \tag{H3}$$

For the resonant case, using a very good approximation for the eigenvectors $A_n^{r,c,j}$ [Eqs. (E14-E15)], the Q function, expressed in terms of the Hypergeometric function, is found to be

$$\widetilde{Q(\alpha,t)} = e^{-|\alpha|^2 - \bar{n}} \sum_{p=0}^{N} \left\{ \left| \sum_{s=0}^{\infty} \frac{|\alpha|^s \bar{n}^{\frac{s}{2}} e^{is\varphi}}{s!} \left(\frac{1}{\sqrt{2}}\right)^{s+p} \right. \right.$$
$$\times \left[ \sum_{j'=0}^{p} e^{ij'\pi} \sqrt{\frac{(s+p)!}{j'!(s+p-j')!}} \sqrt{\frac{(s+p-j')!p!}{j'!s!}} \frac{F(-s,-j',p-j'+1:-1)}{(p-j')!} e^{-i4\pi q_{\frac{N}{2},s-\frac{N}{2}+p,j'}\sqrt{N}\frac{t}{\tau_R}} \right. \tag{H4}$$
$$\left. \left. + \sum_{j'=p+1}^{s+p} e^{ij'\pi} \sqrt{\frac{(s+p)!}{j'!(s+p-j')!}} e^{i(j'-p)\pi} \sqrt{\frac{s!j'!}{(s+p-j')!p!}} \frac{F(j'-s-p,-p,j'-p+1:-1)}{(j'-p)!} e^{-i4\pi q_{\frac{N}{2},s-\frac{N}{2}+p,j'}\sqrt{N}\frac{t}{\tau_R}} \right] \right|^2 \right\}.$$

Similar expressions can be found for the case of a LMPN and small number of TLMs in the up state. Starting from Eq. (H1)

$$Q(\alpha,t) = \langle \alpha|\rho_f(t)|\alpha\rangle$$
$$= e^{-|\alpha|^2} \sum_{s=0}^{\infty} \sum_{r=0}^{\infty} \frac{|\alpha|^s |\alpha|^r}{\sqrt{s!r!}} e^{-i\varphi(r-s)} < s|\rho_f(t)|r>. \tag{H5}$$

For the case of a LMPN and small numbers of TLMs, the inner bracket on $\rho_f(t)$ can be written as

$$\langle n|\rho_f(t)|n'\rangle = \sum_{s=0}^{N} \sum_{j=0}^{N} \sum_{j'=0}^{N} (A^*)_n^{r,n+r-s,j} A_{n-s}^{r,n+r-s,j} e^{-i\lambda_{r,n+r-s,j}t} (A^*)_{n'}^{r,n'+r-s,j'} \tag{H6}$$
$$\times A_{n'-s}^{r,n'+r-s,j'} e^{i\lambda_{r,n'+r-s,j'}t} \langle n-s|\rho_f(0)|n'-s\rangle$$

In Eq. (H6) we replaced s by n and r by n'. To proceed further, assume that the initial photon distribution is given by the coherent distribution $|\beta><\beta|$. Now we know that

$$|\beta> = e^{-\frac{|\beta|^2}{2}} \sum_{q=0}^{\infty} \frac{|\beta|^q e^{-iq\theta}}{\sqrt{q!}} |q> \tag{H7}$$

Thus

$$|\beta><\beta| = e^{-|\beta|^2} \sum_{t=0}^{\infty} \sum_{q=0}^{\infty} \frac{|\beta|^{q+t} e^{-i(q-t)\theta}}{\sqrt{t!\,q!}} |q><t| \tag{H8}$$

The result for $\langle n-s|\rho_f(0)|n'-s\rangle$ is given by

$$\langle n-s|\rho_f(0)|n'-s\rangle = e^{-\bar{n}} \frac{(\bar{n})^{\frac{n+n'-2s}{2}} e^{-i(n-n')\theta}}{\sqrt{(n'-s)!\,(n-s)!}} \tag{H9}$$

This can then be inserted into Eq. (H7) to obtain

$$\langle n|\rho_f(t)|n'\rangle = \sum_{s=0}^{N} \sum_{j=0}^{N} \sum_{j'=0}^{N} (A^*)_n^{r,n+r-s,j} A_{n-s}^{r,n+r-s,j} e^{-i\lambda_{r,n+r-s,j}t}$$

$$\times (A^*)_{n'}^{r,n'+r-s,j'} A_{n'-s}^{r,n'+r-s,j'} e^{i\lambda_{r,n'+r-s,j'}t} e^{-\bar{n}} \frac{(\bar{n})^{\frac{n+n'-2s}{2}} e^{-i(n-n')\theta}}{\sqrt{(n'-s)!\,(n-s)!}} \tag{H10}$$

This can be inserted into Eq. (H6) with some rearrangement of terms to yield:

$$Q(\alpha,t) = e^{-|\alpha|^2-\bar{n}} \sum_{s=0}^{N} \left| \sum_{n=0}^{\infty} \frac{|\alpha|^n (\bar{n})^{\frac{n-s}{2}} e^{-in(\varphi-\theta)}}{\sqrt{n!\,(n-s)!}} \sum_{j=0}^{N} (A^*)_n^{r,n+r-s,j} A_{n-s}^{r,n+r-s,j} e^{-i\lambda_{r,n+r-s,j}t} \right|^2 \tag{H11}$$

Since n has to be greater or equal to s, let p=n-s and Eq. (H11) becomes:

$$Q(\alpha,t) = e^{-|\alpha|^2-\bar{n}} \sum_{s=0}^{N} \left| \sum_{p=0}^{\infty} \frac{|\alpha|^{p+s} (\bar{n})^{\frac{p}{2}} e^{-i(p+s)(\varphi-\theta)}}{\sqrt{(p+s)!\,p!}} \sum_{j=0}^{N} (A^*)_{p+s}^{\frac{N}{2},p+\frac{N}{2},j} A_p^{\frac{N}{2},p+\frac{N}{2},j} e^{-i\lambda_{\frac{N}{2},p+\frac{N}{2},j}t} \right|^2 \tag{H12}$$

If time is set to 0, we use the orthogonality relations for $A_p^{\frac{N}{2},p+\frac{N}{2},j}$ to realize that the only valid value for s is 0. Then

$$Q(\alpha,0) = e^{-|\alpha|^2-\bar{n}} \left| \sum_{p=0}^{\infty} \frac{|\alpha|^p (\bar{n})^{\frac{p}{2}} e^{-ip\psi}}{p!} \right|^2$$

$$= e^{-|\alpha|^2-\bar{n}} e^{2|\alpha|\bar{n}^{\frac{1}{2}}\cos\psi} \tag{H13}$$

For other values of time it is necessary to make an approximation. Note that

$$q = \frac{c - \lambda}{|\kappa|},$$

And for the case were all the TLMs are initially in the upstate.

$$\lambda = p + \frac{N}{2} - |\kappa|q \tag{H14}$$

Thus Eq. (H12) can be written as

$$Q(\alpha, t) = e^{-|\alpha|^2 - \bar{n}} \sum_{s=0}^{N} \left| \sum_{p=0}^{\infty} \frac{|\alpha|^{p+s}(\bar{n})^{\frac{p}{2}} e^{-i(p+s)(\varphi-\theta)}}{\sqrt{(p+s)!\, p!}} e^{-ipt} \sum_{j=0}^{N} (A^*)_{p+s}^{\frac{N}{2}, p+\frac{N}{2}, j} A_p^{\frac{N}{2}, p+\frac{N}{2}, j} e^{iq_{\frac{N}{2}, p+\frac{N}{2}, j} \Omega |\kappa| t} \right|^2 \tag{H15}$$

Note that the N term in Eq. (H14) can be ignored since it is a constant and doing the absolute square in Eq. (H15) makes it disappear. The approximation is now to ignore the term $e^{-ipt}$ in Eq. (H15). The reason being that this term oscillates very quickly compared to the q term and we could choose times to perform the calculation such that that term would be unity. Further the expression only has significant value only for p near $\bar{n}$. Using normalized time Eq. (H15) becomes:

$$Q(\alpha, t) = e^{-|\alpha|^2 - \bar{n}} \sum_{s=0}^{N} \left| \sum_{p=0}^{\infty} \frac{|\alpha|^{p+s}(\bar{n})^{\frac{p}{2}} e^{-i(p+s)(\varphi-\theta)}}{\sqrt{(p+s)!\, p!}} \sum_{j=0}^{N} (A^*)_{p+s}^{\frac{N}{2}, p+\frac{N}{2}, j} A_p^{\frac{N}{2}, p+\frac{N}{2}, j} e^{2\pi i \sqrt{\bar{n}+1+\Delta} q_{\frac{N}{2}, p+\frac{N}{2}, j} \tau} \right|^2. \tag{H16}$$

This is the exact results. It is possible to obtain approximate results using the average field approach and in particular the expressions found in Eq. (D16).

Namely for the term $A_p^{\frac{N}{2}, p+\frac{N}{2}, j}$

$$A_p^{\frac{N}{2}, p+\frac{N}{2}, j} = (b1)^j (a1)^{N-j} \sqrt{\frac{(N)!}{j!\,(N-j)!}}. \tag{H17}$$

For the resonant case ( the only case we have performed calculations, Eq. (H17) becomes

$$A_p^{\frac{N}{2}, p+\frac{N}{2}, j} = \frac{1}{2^{\frac{N}{2}}} \sqrt{\frac{(N)!}{j!\,(N-j)!}}. \tag{H18}$$

Again using Eq. (D16)

$$A_{p+s}^{(\frac{N}{2}, n+\frac{N}{2}, j)} = (b1)^j (a1)^{N-j} \begin{cases} \left(\frac{a2}{a1}\right)^s \sqrt{\frac{(N-j)!\,(N-s)!}{s!\,j!}} \dfrac{F\left(-j,-s,N+1-s-j;-\left(\frac{b2a1}{b1a2}\right)\right)}{(N-j-s)!} & N \geq j+s \\[2ex] (-1)^{s-N} \left(\frac{b2a2}{b1a1}\right)^s \left(\frac{b2a1}{b1a2}\right)^{-N} \sqrt{\frac{s!\,j!}{(N-s)!\,(N-j)!}} \dfrac{F\left(j-N, s-N, 1+s+j-N;-\left(\frac{b2a1}{b1a2}\right)\right)}{(j+s-N)!} & N < j+s \end{cases} \tag{H19}$$

$$A^{(\frac{N}{2},n+\frac{N}{2},j)}_{p+s} = \frac{1}{2^{\frac{N}{2}}}\begin{Bmatrix} \sqrt{\frac{(N-j)!(N-s)!}{s!j!}}\frac{F(-j,-s,N+1-s-j:-1)}{(N-j-s)!} & N \geq j+s \\ (-1)^{s-N}\sqrt{\frac{s!j!}{(N-s)!(N-j)!}}\frac{F(j-N,s-N,1+s+j-N:-1)}{(j+s-N)!} & N < j+s \end{Bmatrix} \quad \text{(H20)}$$

Equation (H20) is for the resonant case. Inserting Eqs. (H18) and (H20) into Eq. (H16) for the resonant case one obtains the approximate solution

$$Q(\alpha,t) = \frac{1}{2^N}e^{-|\alpha|^2-\bar{n}}\sum_{s=0}^{N}\left|\sum_{p=0}^{\infty}\frac{|\alpha|^{p+s}(\bar{n})^{\frac{p}{2}}e^{-i(p+s)(\varphi-\theta)}}{\sqrt{(p+s)!p!}}\right.$$
$$\times\left[\sum_{j=0}^{N-s}\sqrt{\frac{(N)!}{j!(N-j)!}}\sqrt{\frac{(N-j)!(N-s)!}{s!j!}}\frac{F(-j,-s,N+1-s-j:-1)}{(N-j-s)!}e^{2\pi i\sqrt{\bar{n}+1+\Delta}q_{\frac{N}{2},p+\frac{N}{2},j}\tau}\right.$$
$$\left.\left.+\sum_{j=N-s+1}^{N}(-1)^{s-N}\sqrt{\frac{(N)!}{j!(N-j)!}}\sqrt{\frac{s!j!}{(N-s)!(N-j)!}}\frac{F(j-N,s-N,1+s+j-N:-1)}{(j+s-N)!}e^{2\pi i\sqrt{\bar{n}+1+\Delta}q_{\frac{N}{2},p+\frac{N}{2},j}\tau}\right]\right|^2 \quad \text{(H21)}$$

This completes the mathematical background.